%% file: main.tex
\documentclass[a4paper,fleqn,usenatbib]{mnras}
\RequirePackage{fixltx2e}

\usepackage{txfonts}
\usepackage{enumerate}
\usepackage{hyperref}

\usepackage[T1]{fontenc}
\usepackage{ae,aecompl}
\usepackage{multicol}

\usepackage{graphicx}	
\usepackage{amssymb}	
\usepackage{wasysym}
\usepackage{bm}          
\usepackage{color}
\usepackage{xspace}
\usepackage{ulem}

\newcommand{\text}[1]{\mathrm{#1}}

\newcommand{\Msun}{\ensuremath{\,\mathrm{M}_\odot}\xspace}
\newcommand{\Rsun}{\ensuremath{\,\mathrm{R}_\odot}\xspace}
\newcommand{\teff}{\ensuremath{\,T_\mathrm{\!eff}}\xspace}
\newcommand{\sk}{Sk--$69\,^{\circ}202$\xspace}

\newcommand{\kms}{{\ensuremath{\,\mathrm{km}\,\mathrm{s}^{-1}}}\xspace}

\newcommand{\B}{{\ensuremath{\,\mathrm{B}}}\xspace}

\title[Explosions of binary merger progenitors, SN~1987A]{Explosions of blue supergiants from binary mergers for SN~1987A}

\author[Menon, Utrobin, \& Heger]{
Athira Menon,$^{1,2}$
Victor Utrobin,$^{3,4}$
Alexander Heger,$^{1,2,5}$
\\
$^{1}$School of Physics and Astronomy, Monash University, Clayton, VIC 3800, Australia\\
$^{2}$Monash Centre for Astrophysics\\
$^{3}$Max-Planck-Institut f\"{u}r Astrophysik, Karl-Schwarzschild-Str. 1, 85748 Garching, Germany\\
$^{4}$State Scientific Center of the Russian Federation, Institute for Theoretical and Experimental Physics of National Research Center, \\
``Kurchatov Institute", B. Cheremushkinskaya St. 25, 117218 Moscow, Russia \\
$^{5}$Tsung-Dao Lee Institute, Shanghai 200240, China\\
}

\date{}

\pubyear{2017}

\begin{document}
\label{firstpage}
\pagerange{\pageref{firstpage}--\pageref{lastpage}}
\maketitle

\begin{abstract}
Based on the work of Menon and Heger (2017), we present the bolometric light curves and spectra of the explosions of blue supergiant progenitors from binary mergers.  We study SN~1987A and two other peculiar Type IIP supernovae: SN~1998A and SN~2006V.  The progenitor models were produced using the stellar evolution code {\sc Kepler} and then exploded using the 1D radiation hydrodynamic code {\sc Crab}.  The explosions of binary merger models exhibit an overall better fit to the light curve of SN~1987A than previous single star models, because of their lower helium-core masses, larger envelope masses, and smaller radii.  The merger model that best matches the observational constraints of the progenitor of SN~1987A and the light curve is a model with a radius of $37\,\Rsun$, an ejecta mass of $20.6\,\Msun$, an explosion energy of  $1.7\times10^{51}\,$erg, a nickel mass of $0.073\,\Msun$, and a nickel mixing velocity of $3,\!000\,\kms$.  This model also works for SN~1998A and is comparable with earlier estimates from semi-analytic models.  In the case of SN~2006V, however, a model with a radius of $150\,\Rsun$ and ejecta mass of $19.1\,\Msun$ matches the light curve.  These parameters are significantly higher than predictions from semi-analytic models for the progenitor of this supernova.  
\end{abstract}


\input{Introduction.tex}
\input{Methodology.tex}
\input{Results.tex}
\input{Conclusions.tex}
\section*{Acknowledgements}
We thank Alexandra Kozyrova and Luc Dessart for useful initial discussions on this project.  We also thank Andrea Pastorello for the data on SN~1998A and Francesco Taddia for the data on SN~2006V.  We also thank Natasha Ivanova and Philipp Podsiadlowski for initial  discussions about the progenitor model and the anonymous referee for his encouraging comments and suggestions.  AH was supported by an ARC Future Fellowship (FT120100363). 

\clearpage
\bibliographystyle{mnras}
\bibliography{master} 
\bsp	
\label{lastpage}
\end{document}

%% file: Introduction.tex
\section{Introduction}
\label{intro}

Type II plateau supernovae (Type IIP SNe) are the most common class of    core-collapse supernovae whose progenitors are red supergiants (RSGs) of    a few hundred solar radii with large hydrogen-rich envelopes (some    $\gtrsim10\,\Msun$) and contains an iron core at the time of collapse    \citep{GIN_71,falk1977,smartt2009,dessart2010a}.  They are characterized by a distinct plateau phase in their bolometric luminosity evolution, typically of $80-100$\,days, followed by the nebular phase where the light curve descends to a long tail powered by radioactive decay of $^{56}\mathrm{Co}$ to $^{56}\mathrm{Fe}$.

In February 1987, however, a new H-rich Type II SN was spotted, in the Large Magellanic Cloud (LMC).  Although the spectrum of SN~1987A contained H-lines, its light curve was dome-shaped, with a much slower rise to maximum compared to Type IIP SNe and a near absence of a plateau.  These features indicated that the progenitor was far more compact than a RSG.  Revisiting the pre-explosion images of the LMC showed that the progenitor was indeed a hot, compact, blue supergiant \citep[BSG;][]{walborn1987}.  \sk had a luminosity of $\log\,L/\mathrm{L}_{\odot}=4.9-5.1$, an effective temperature of $\teff=15-18\,$kK, and a radius of $R\leq50\Rsun$ \citep{woosley1988a,barkat1989a}.  The small progenitor radius helped explain both its slowly rising light curve  ($\approx84\,$days) and its low peak luminosity compared to typical Type IIP SNe \citep{wood1988,arnett1989}.  \sk is by far the most well-observed of all supernovae progenitors.  A curious circumstellar nebular structure, in the form of a triple-ring, was observed around the central debris site \citep{burrows1995}, which was ejected by the parent star, at least 20,000 years before exploding \citep{burrows1995, sugerman2005a}.  The rings themselves were enriched in nitrogen compared to carbon and oxygen, with number fraction ratios of $\text{N/C}\sim5\pm2$ and $\text{N/O}\sim1.1\pm0.4$    \citep{lundqvist1996}, and in helium compared to hydrogen with a number fraction ratio of $\text{He/H}= 0.14\pm0.06$ \citep{france2011}.  These abundances originate from the surface of the progenitor during its phase as a BSG and indicate enrichment by material that underwent CNO burning at some time during the evolution of the projenitor.

The first evolutionary models that attempted to reproduce the observational features of \sk were from single stars.  These models were derived by fine-tuning specific aspects of the evolutionary path to make the star explode as a BSG, such as rotation rate    \citep{weiss1988}, reducing the helium-core mass \citep{barkat1989b}, decreasing the envelope metallicity \citep{weiss1988}, increasing the abundance of helium in the envelope, increasing mass loss \citep{saio1988, shigeyama1990}, restricting the convection    \citep{woosley1988a}, or inducing rotation and restricting semi-convection \citep{woosley1997}.  Along with the extreme physics evoked, except for the spin-up mechanism described by \citet{heger1998}, these single star models could not explain how the progenitor could gain enough angular momentum to eject material that would form the complex triple-ring nebular structure    \citep{pods1992b,smartt2009}.  The first evolutionary calculations that explored the role of accretion or a merger in the binary system, were done by \citet{pods1989} and \citet{pods1990}, in which two of the models evolved into BSGs that tallied with the location of \sk.  The merger evolutionary scenario was further developed in \citet{pods1992a,pods2006} and the 3D hydrodynamic simulations of the merger itself were studied in \citep{ivanova2002b,ivanova2002c,ivanova2003}.  The 3D simulations of \citet{morris2009} also demonstrated that the triple-ring nebular structure could be reproduced by using a merger model.

The light curve of SN~1987A can be classified into three
parts: the early light curve up to $20-30\,$days was powered by the release of the internal energy left behind by the shock wave after propelling the expansion of the ejecta; the middle part of the light curve was entirely powered by the radioactive decay energy of $^{56}\mathrm{Ni}$ to $^{56}\mathrm{Co}$; and the late part of the light curve, $>100$\,days, was powered by the decay of $^{56}\mathrm{Co}$ to $^{56}\mathrm{Fe}$ \citep{woosley1988b,shigeyama1990,utrobin2004}.  These features of the light curve of SN~1987A make it significantly different from those of typical Type IIP SNe \citep{woosley1988b,arnett1989,hamuy2003}.

The fast and smooth rise of the light curve and its broad dome shape indicated a large extent of mixing of $^{56}$Ni out into the envelope where it could not have been produced by thermonuclear reactions, along with the mixing of H and other chemical species into the He core \citep{woosley1988b,shigeyama1990,blinnikov2000,utrobin2004}.  Spectral observations between $20-100\,$days showed an unusual phenomenon called the \textsl{Bochum event}, wherein fine structures were observed in the $\mathrm{H}{\alpha}$ line profile \citep{hanuschik1988,phillips1989}.  In addition, the net flux of the $\mathrm{H}{\alpha}$ line dropped to zero at Day 20, but was greater than zero before and after that, which indicated that an additional heating source powered the $\mathrm{H}{\alpha}$ line after its drop at Day 20 \citep{phillips1989,thimm1989}.  This heating source was attributed to the emission of $\gamma$-rays from the decay of $^{56}$Ni and $^{56}$Co clumps or fingers, which were mixed from the explosion engine to the outer layers of the ejecta where H was present \citep{woosley1998}.  The mixing is expected to occur when the shock wave from the explosion gets decelerated at the (C+O)/He and He/H composition interfaces.  The formation of Rayleigh-Taylor (RT) instabilities at these interfaces causes outward mixing of $^{56}\mathrm{Ni}$ into the hydrogen envelope, whereas inward mixing of hydrogen into the helium core of the progenitor star only depends on the strength of RT instabilities at the He/H composition interface (see, e.g., \citealp{arnett1989, shigeyama1990, kifonidis2003, kifonidis2006, wong2015}).

Observations also supported the requirement of strong mixing of species in the ejecta: the bulk of the Ni and Fe mass in the ejecta was observed to mix out to regions travelling at $3,\!000\,$km/s \citep{colgan1994}. \citet{utrobin1995} could reproduce the Bochum event by accounting for the fast, outward mixing of Ni clumps with one clump of Ni found to be travelling as fast as $4,\!700\,$km/s.  The later evolution ($>200$\,days) of velocity estimated from the $\mathrm{H}{\alpha}$ line profile in the ejecta showed that H was distributed down to regions within the He core, where the ejecta had velocities of $500-700\,$km/s \citep{chugai1991, kozma1998, MJS_12}.  Observations of the 3D distribution of the ejecta by \citet{larsson2016} found H$\alpha$ line emission in the ejecta down to velocities of $450\,$km/s.

Since its discovery, eleven more supernovae have been identified with very similar spectra and light curve shapes as SN~1987A \citep{pastorello2012,taddia2013}.  Together, they are classified as Type II-peculiar supernovae (Type II-pec SNe), and they are all expected to have massive BSG progenitors.  The prototype of this class is SN~1987A, which to this date remains the most well-recorded supernova with data from as early as 1 day until it faded away a year later.

Supernova explosion studies can be classified in two categories: those based on evolutionary models and those on non-evolutionary models.  The first one uses a progenitor model constructed from stellar evolution    calculations and then varies the explosion parameters to produce a match with the observations.  The second one is a reverse-engineering process, in which one constructs an `optimal' stellar model by assuming a homologous relation between mass and radius.  In this method, individual parameters such as the radius, ejecta mass, explosion energy, and mixing are varied to obtain the best parameter set for reproducing the light curve of the supernova.

We use the following notations in this paper: $R_\mathrm{pre-SN}$ is the radius of the progenitor (or pre-SN) model and $M_\mathrm{pre-SN}$ is the mass of the pre-SN model. $M_\mathrm{ej}$ is the ejecta mass obtained from $M_\mathrm{ej}= M_\mathrm{pre-SN}-M_\mathrm{cut}$, where $M_\mathrm{cut}$ is the baryonic mass cut that determines the neutron star mass and is the inner mass boundary of the progenitor which is not included in the explosion (see Section~\ref{sec:models-lcurves}). $E$ is the explosion energy applied at the mass cut, and for which we use the unit Bethe (B) where $1\,$B$=10^{51}\,$ergs; it is the excess above the total energy of the envelope beyond the mass cut and hence is essentially converted into the kinetic energy of the ejected mass.  The $^{56}\mathrm{Ni}$ mass in the ejecta is denoted as $M_\mathrm{Ni}$ and the velocity of the ejected $^{56}\mathrm{Ni}$ mass is denoted as $v_\mathrm{Ni}$.

Explosion studies that have used these single star evolutionary models have been done by, e.g., \citet{woosley1988b, nomoto1988, saio1988,arnett1989, shigeyama1990, blinnikov2000, utrobin2004}, using 1D radiation hydrodynamic codes.  These models have a radius of $R_\mathrm{pre-SN}=30\,\Rsun-70\,\Rsun$, He core mass of $M_\mathrm{He,c}=4\,\Msun-7.4\,\Msun$ and an envelope mass of $M_\mathrm{env}=9\,\Msun-14\,\Msun$.  With a choice of explosion parameters that include $E=0.6\,\B-1.65\,\B$,  $M_\mathrm{Ni}\approx0.07\,\Msun$, which was determined from the luminosity of the nickel decay tail of the light curve, and a `strong' nickel mixing velocity compared to the observed value of $3,\!000\,$\kms of $v_\mathrm{Ni}=4,\!000\,$\kms, these models produced reasonably good fits with the light curve shape.  \citet{dessart2010b} used a more sophisticated 1D time-dependent radiative transfer code to study the first 20 days of the spectral and light curve evolution, with an $18\,\Msun$ progenitor model which had $M_\mathrm{ej}=15.4\,\Msun$ and $R_\mathrm{pre-SN}=47\,\Rsun$.  The results from the explosion of this model also matched the observations in this period very well.

Non-evolutionary explosion calculations carried out by  \citet{utrobin1993, utrobin2004, utrobin2005} predicted different results for the progenitor than the above evolutionary calculations.  \citet{utrobin1993} demonstrated that a progenitor model with $M_\mathrm{ej}=15\,\Msun-19\,\Msun$ and a density structure with a polytropic index of $n=3$, when exploded with $E=1.25\,\B-1.65\,\B$, can fit the light curve.  \citet{utrobin2004} scaled an evolutionary model which had $M_\mathrm{ej}=17.8\,\Msun$ and $R_\mathrm{pre-SN}=64.2\,\Rsun$ to one with $M_\mathrm{ej}=18\,\Msun$ and $R_\mathrm{pre-SN}=46.8\,\Rsun$.  By using an explosion energy of $1\,\B$ and a nickel mass of $0.073\,\Msun$ with this scaled model, \citet{utrobin2004} could produce an excellent fit to the light curve data, but again with a strong nickel mixing velocity of $4,\!000\,\kms$.  On the other hand, an artificially constructed  progenitor model which had $M_\mathrm{ej}=18\,\Msun$, $R_\mathrm{pre-SN}=46.8\,\Rsun$ and a modified density distribution compared to above evolutionary models, could reproduce the observed light curve with a `moderate' nickel mixing velocity of $2,\!500\,\kms$, which is much closer to the observed value.

Except for \citet{dessart2010b}, all the above studies, both evolutionary and non-evolutionary, used only the photometric data of SN~1987A to constrain the progenitor model and did not study the evolution of the spectral lines in the ejecta.  Using a non-evolutionary model of \citet{utrobin2004}, \citet{utrobin2005} were the first to simultaneously constrain a progenitor model by including both the photometric and spectroscopic observations of SN~1987A.   Based on this study, \citet{utrobin2005} predicted that the optimal progenitor model whose explosion can match the bolometric light curve, the kinetics of spectral lines and the Ni mixing velocity of $v_\mathrm{Ni}=3,\!000\,\kms$ required from observations, has $R_\mathrm{pre-SN}=35\pm5\,\Rsun$, $M_\mathrm{ej}=18\pm1.5\,\Msun$, and an explosion energy to ejecta mass ratio of $E/M_\mathrm{ej}=0.83\times10^{50}\,$ergs/\Msun, which gives $E=1.5\pm0.12\,$B for this model.

The 3D simulations of \citet{wong2015} demonstrated that the extent of mixing    in the supernova explosion depends heavily on the overall progenitor density    structure, in particularly the density gradient at the He/H interface, the compactness of the CO core, and the location of the composition interfaces.  It was found that the nickel fingers penetrate deeper in RSG models, between    $4,\!000-5,\!000\,$\kms in the H-rich envelope, than in BSG models in which the fingers penetrated out only up to velocities of $2,\!200\,\kms$, except in model B15 in which nickel was mixed out to a maximum velocity of $3,\!500\,\kms$.  The growth rate of Rayleigh-Taylor (RT) instabilities depends on the location where they are formed, which is connected to the density structure of the He core and the formation time of the reverse shock that forms when the shock wave is decelerated as it passes through the He/H interface.  For BSGs, nickel plumes in the He core do not have sufficient time to grow and penetrate the envelope before being stalled by the reverse shock.  In order to allow the growth of nickel fingers, a steeper density gradient at the He/H interface, as typically found in RSGs, is required in the BSG models as well.

The most detailed work so far that did a complete study of the evolution of the supernova ejecta from the early explosion phase until the late nebular phase is that of \citet{utrobin2015}.  Using certain single-star BSG progenitor models, they performed 3D neutrino-driven explosion simulations to study the evolution of hydrodynamic and chemical composition quantities until shock breakout and thereafter mapped these quantities to a 1D spherically symmetric model to obtain the light curve shape and the evolution of the photospheric velocity profiles.  None of the single-star evolutionary models used in the study could reproduce the light curve shape or spectral observations of SN~1987A, except for the optimal model of \citep{utrobin2005}.  The observed nickel mixing velocity of $3,\!000\,\kms$ was found only in one of the evolutionary models (B15), with $M_\mathrm{ej}=15\,\Msun$ and $R=56.1\,\Rsun$, whereas the other evolutionary models only produced nickel mixing velocities up to $2,\!000\,$\kms.  Hydrogen was found to be mixed inward down to velocities of $100\,\kms$ in all models, in agreement with observations.  The mixing of hydrogen determines the shape of the broad dome of the light curve.

Thus there are two reasons why we need new pre-SN evolutionary models for SN~1987A.  First, despite strong proof that the progenitor of SN~1987A evolved from a binary merger, there are no pre-SN models in published literature based on the binary merger scenario that match both, the observed signatures of \sk and whose explosions also match the light curve and spectral features of the supernova.  Second, there is a need for an evolutionary model whose 3D neutrino-driven explosion can reproduce the required penetration of nickel and hydrogen fingers in agreement with the observed velocities of mixed matter and thereafter fit the light curve as well.  \citet{utrobin2015} already predicted that a progenitor from a binary merger can have larger envelope mass and may be more favourable to match the overall light curve shape of SN~1987A.

All explosion studies for SN~1987A so far have used single star pre-SN models, which were derived by fine-tuning specific aspects of the evolution to make the star explode as a BSG.  \citet[][hereafter Paper~I]{menon2017} conducted the first systematic and detailed study of binary merger evolutionary models based on the merger scenario of \citet{pods1992a, pods2007}, which were evolved until the pre-SN stage, i.e., until just prior to the onset of iron-core collapse.  Six of the $84$ pre-SN models matched the observational criteria of \sk and the majority of the pre-SN models were blue.  The study included a large range of initial parameters, including the primary and secondary masses and mixing boundaries during the merging.  In this paper, we use these pre-SN models to conduct the first study of explosions of binary merger progenitors for SN~1987A.  We broaden our study to investigate the viability of our progenitor models for two other peculiar Type IIP SNe: SN~1998A and SN~2006V.  We would especially like to thank the authors of \citet{lusk2017} for putting together the bolometric light curve data and nickel masses of five SN~1987A-like objects.

The merger pre-SN models have been computed with code {\sc Kepler} and the light curves were calculated using code {\sc Crab}, as explained in Section~\ref{Methodology}.  In Section~\ref{Results} we explain the role of different parameters affecting the shape of the light curve, how they compare with earlier single star models and the best fit model for SN~1987A.  We also assess whether the merger models can explain other peculiar Type IIP supernovae and thus infer the properties of their progenitors.

%% file: Methodology.tex
\section{Methodology}
\label{Methodology}
\subsection{{\sc Kepler}: the stellar evolution code }
In Paper I, we constructed post-merger models using the implicit hydrodynamic stellar evolution code {\sc Kepler}  \citep{woosley2002, woosley2007a}.
The code assumes a dynamo model \citep{spruit2002} and includes the physics for angular momentum
   transport due to rotation \citet{heger2000a,heger2000b}, and  prescriptions for mixing arising from rotationally-induced instabilities \citep{heger2000a, heger2005}. Up to oxygen depletion in the core, energy generation treated using a 19-isotope network that comprises the key reaction needed, and after oxygen depletion the code switches to a 128 isotope nuclear quasi-statistical equilibrium model, and after silicon burning to a nuclear statistical equilibrium network \citep{heger2010, sukhbold2016}. The reaction rates used are as described in \citet{rauscher2000,woosley2007a}. 

In Paper I, we also updated the opacity routines in the code, to include Type II OPAL opacities, conductive and molecular opacities using the routines built by \citet{consta2014}. Mass loss is implemented using to the prescription by \citet{dejager1988}, but scaled with metallicity as $\left(Z/Z_\odot\right)^{0.5}$.

\subsection{The binary merger}

\subsubsection{Evolution leading up to the binary merger} 
\label{pods_merger}
The progenitor models of Paper I were built according to the binary merger scenario outlined in \citet{pods2007}, and incorporating results from merger simulations of \citet{ivanova2002a} and from 1D post-merger simulations of \citet{ivanova2003}.
The evolution of the binary sequence is as follows \citep{pods1992a,pods1992b, pods2007}: A massive main-sequence primary (between $15\,\Msun-20\,\Msun$) and an intermediate mass main-sequence secondary ($\approx5\,\Msun$) have an initial orbital period of $\sim10\,$yr. When the primary star evolves to a red giant with a He-depleted core, it overflows its Roche Lobe and transfers mass on a dynamically unstable timescale to the secondary, which leads to a common envelope episode. The secondary and the core of the primary spiral in towards each other as they lose energy due to frictional forces. This loss of orbital energy, causes a partial ejection of the envelope. The spiral-in phase continues until the secondary also fills its Roche Lobe and transfers mass on the He core of the primary, over a period of the order of a 100\,yr \citep{ivanova2002b,ivanova2002c}. At the end of the mass-transfer phase, the secondary is expected to be completely disrupted and dissolves in the envelope of the primary. In the hydrodynamic simulations of \citet{ivanova2002c},  a stream of the H-rich secondary mass was found to penetrate the He core during the mass transfer and dredge up He-rich material to the surface, thus causing the He core to shrink in mass. After the merging phase is completed, the thermally unstable RSG structure is expected to first puff up and then contract to a BSG.  The star reaches core collapse as a BSG and explodes as a Type II SNe  \citep{ivanova2003}.  The present models do not consider spin-up by the merger.  \citet{ivanova2016} find, although for low-mass common-envelope mergers, that most of the angular momentum is quickly lost along with the ejecta from the surface.  

It should be noted that there are alternative binary scenarios to the one used in this work to produce BSG pre-SN models. These involve a primary that undergoes a Case B mass transfer when the primary is still burning He in its core, as discussed in \citet{pods1989,pods1992a,pods1992b}.  A BSG may form either from the merger of a secondary main-sequence star with the primary, or the secondary could itself evolve to a BSG after the accretion of mass from the primary.

\subsubsection{Blue supergiants from post-binary mergers (Paper I)}  
In Paper I, evolutionary models were constructed by considering three initial parameters: the primary mass,  $M_{1}$, the secondary mass, $M_{2}$, and the fraction of the He shell of the primary's He core dredged up, $f_\mathrm{sh}$. The range of these parameters were:  $M_{1}=15\,\Msun$, $16\,\Msun$, and $17\,\Msun$;  $M_{2}=2\,\Msun$, $3\,\Msun$, $\ldots$, $8\,\Msun$ and $f_\mathrm{sh}=10\,\%$, $50\,\%$, $90\,\%$, and $100\%$, where $100\,\%$ corresponds to the boundary of the CO core.

The evolution was followed from the zero age main sequence of the rotating primary star ($v/v_\mathrm{crit}=30\,\%$), until the central helium abundance dropped to $\sim10^{-2}$, i.e., until the end of core He-burning. At this stage the primary is an RSG with a He-depleted core and it merges with the secondary main-sequence star
   according to the scenario of \citet{pods2007}.
A simple 1D merging prescription was implemented, based on the merger phenomenon described in Section~\ref{pods_merger}. At the end of the merger the He-core mass becomes smaller, depending on
   the mixing boundary set by $f_\mathrm{sh}$, and the envelope mass becomes larger, depending on the mass of the secondary that was accreted. The model was evolved until just prior to the onset of iron core-collapse, when the infall velocity in any part of the star exceeded $\sim1,\!000\,\,$km$\,$s$^{-1}$. This sets the pre-supernova (pre-SN) model.

Two important factors that determined the final position of the
   post-merger star in the HR diagram were $f_\mathrm{sh}$ and
   $M_{2}$. In general, increasing $f_\mathrm{sh}$ for a given combination of $M_{1}$ and $M_{2}$
   led to smaller He cores and resulted in the pre-SN model to have a lower $T_\mathrm{eff}$,
   whereas increasing $M_{2}$ for given $M_{1}$ and $f_\mathrm{sh}$ caused the pre-SN model to have a higher  $T_\mathrm{eff}$.
For the parameter space we considered, BSG pre-SN models,
   which have $\teff \geq 12\,\text{kK}$, always formed when
   $f_\mathrm{sh}=10\,\%$ to $50\,\%$ and $M_{2}\geq4\,\Msun$.
In the study of Paper I, $56$ of the $84$ pre-SN models computed were BSGs whereas the rest were yellow supergiants (YSGs), with $7\leq \teff < 12\,\text{kK}$. 

\subsection{Progenitor models used in this work}

Although eleven peculiar Type IIP SNe are known \citep{pastorello2012}, only five of these have sufficiently good observational data \citep{lusk2017} aside
   from SN~1987A.
These are: SN~1998A, SN~2000cb, SN~2006au, SN~2006V, and SN~2009e.
SN~2000cb was studied by \citet{utrobin2011} and its progenitor was expected
   to have a radius of $35\pm14\,\Rsun$ and an ejecta mass of $22.3\pm1\,\Msun$.
We attempted to make new merger models with the ejecta mass of
   $22.3\pm1\,\Msun$, however, their radius was much larger than
   $35\pm14\,\Rsun$.
Hence we could not study SN~2000cb in this work.
For SN~2006au, there were no data points for the cobalt decay tail luminosity
   in the bolometric light curve and therefore we could not investigate this
   supernova in this study.
In the case of SN~2009e, there were not enough data points in the bolometric
   light curve for $t<100\,$days \citep{pastorello2012}, without which it was
   not possible to perform a progenitor analysis.
Hence SN~2009E was also omitted from our study.

In this work, we use the pre-SN models of Paper I for SN~1987A and SN~1998A and computed additional models for SN~2006V.  These are all listed in Table~\ref{presn2}. The progenitor models for SN~1987A
   are those that satisfy all the observational criteria of \sk, viz., the N/C and N/O ratios in its surface, and its position in the HR diagram. We expand the range of models we investigate for SN~1987A to study the impact of changing the parameters $f_\mathrm{sh}$ and $M_{2}$ on the explosion
   results. 
   
\begin{table*}
 \caption{\textsl{Model} denotes the name of the pre-SN model; $M_\text{1}$ and
    $M_\text{2}$ are the initial primary and secondary masses of the binary
    system; $f_\text{sh}$ is the fraction of the He-shell mass dredged up;
    $M_\text{cut}$ is the mass cut determined at the specific entropy of
    4\,$k_\text{b}$\,baryon$^{-1}$;
    $M_\text{He\,c}$, $M_\text{Fe\,c}$, $M_\text{env}$, and $M_\text{pre-SN}$
    are He-core, iron-core, envelope masses and mass of the pre-SN model
    ($M_\text{He,c}+M_\text{env}$); $\teff$, $\log(L)$ and $R_\text{pre-SN}$
    are the effective temperature, luminosity, and radius of pre-SN model and
    $\xi_{1.5}$ is the compactness parameters for $M=1.5\,\Msun$.}
 \label{presn2}

 \begin{tabular}{lrrrrrrrrrrrrrrr}
  \hline
Model & $M_\text{1}$ & $M_\text{2}$ & $f_\text{sh}$ & $M_\text{cut}$ & $M_\text{He\,c}$ & $M_\text{env}$ & $M_\text{pre-SN}$ & $M_\text{Fe\,c}$ & $T_\text{eff}$ & $\log\left(L\right)$  & $R_\text{pre-SN}$ & $\xi_{1.5}$ \\
 & (\Msun) & (\Msun) & (\%) & (\Msun) & (\Msun) & (\Msun) & (\Msun) &  (\Msun)  &(kK) & ($\text{L}_{\odot}$) & ($\text{R}_{\odot}$) &  \\
  \hline
\bf{SN~1987A} & & & & & & & & & \\
  \hline
16-5a & 16 & 5 & 10 & 1.49 & 4.0 & 16.0 & 20.0 & 1.52 & 16.8 & 4.9 & 32.8 & 0.011 \\
16-6a & 16 & 6 & 10 & 1.52 & 4.0 & 16.0 & 20.0 & 1.52 & 16.8 & 4.9 & 32.8 & 0.008 \\
16-6b & 16 & 6 & 50 & 1.59 & 3.6 & 17.4 & 21.0 & 1.50 & 16.9 & 4.9 & 35.7 & 0.009 \\
16-6d & 16 & 6 & 90 & 1.38 & 3.1 & 17.9 & 21.0 & 1.40 & 12.8 & 4.7 & 46.0 & 0.006 \\
16-6c & 16 & 6 & 100 & 1.49 & 3.1 & 17.9 & 21.0 & 1.38 & 11.0 & 4.8 & 64.3 & 0.008 \\
16-7a & 16 & 7 & 10 & 1.65 & 3.8 & 18.2 & 22.0 & 1.53 & 16.9 & 4.9 & 30.8 & 0.009 \\
16-7b & 16 & 7 & 50 & 1.55 & 3.4 & 18.6 & 22.0 & 1.38 & 15.8 & 4.9 & 37.4 & 0.009 \\
16-8a & 16 & 8 & 10 & 1.65 & 3.8 & 19.2 & 23.0 & 1.53 & 18.8 & 5.0 & 28.8 & 0.009 \\
\hline
\bf{SN~1998A} & & & & & & & & & \\
\hline
16-7b & 16 & 7 & 50 & 1.55 & 3.4 & 18.6 & 22.0 & 1.38 & 15.8 & 4.9 & 37.4 & 0.009 \\
\hline
\bf{SN~2006V} & & & & & & & & & \\
\hline
18-4d & 18 & 4 & 90 & 1.64 & 3.8 & 16.7 & 20.5 & 1.50 & 7.5 & 4.8 & 150.3 & 0.009 \\
  \hline
 \end{tabular}
\end{table*}

\subsection{{\sc Crab}: The explosion code}
\label{sec:models-lcurves}
%
The implicit Lagrangian radiation hydrodynamics code {\sc Crab}
   \citep{utrobin2004, Utr_07} integrates the spherically symmetric equations.
It solves the set of hydrodynamic equations including self-gravity, and
   a radiation transfer equation.
The latter is treated in the one-group (gray) approximation in the outer,
   optically transparent or semitransparent layers of the SN ejecta and
   is as the diffusion of equilibrium radiation in the approximation
   of radiative heat conduction in the inner, optically thick layers, where
   thermalization of radiation takes place \citep[e.g.,][]{MM_84}.
The pre-SN models provided by the evolutionary simulations of binary mergers are used as the initial data in our hydrodynamic modelling of the SN outburst. The SN explosion is approximated by the collapse of the inner  core, which is removed from the computational mass domain and collapses into a neutron star. The explosion is initiated by a supersonic piston applied to the location where the entropy drops to S=4\,kb/baryon (i.e., $M_\mathrm{cut}$) near the bottom of the stellar envelope and by assuming no fallback mass.

The time-dependent radiative transfer equation is written in a co-moving frame of
   reference to accuracy of order $v/c$ ($v$ is the fluid velocity,
   $c$ is the speed of light) and is solved for the zeroth and first angular
   moments of the non-equilibrium radiation intensity.
To close the system of moment equations, a variable Eddington factor is
   evaluated, directly taking the scattering of radiation in the ejecta into
   account.
The total set of equations is discretized spatially on the basis of the method
   of lines \citep[e.g.,][]{HNW_93, HW_96}.
The resultant system of ordinary differential equations is integrated using
   the implicit method of \citet{Gea_71} with an automatic choice of both
   the time integration step and the order of accuracy of the method.
Shock waves are automatically captured by means of the linear and nonlinear
   artificial viscosity of \citet{CSW_98}.
In addition, Compton cooling and heating is included in the radiation
   hydrodynamic equations and treated according to \citet{Wey_66}.
Moreover, the calculation of the SN bolometric luminosity allows for
   retardation and limb-darkening effects.

Energy deposition of gamma rays from the decay chain $^{56}$Ni $\to ^{56}$Co
   $\to ^{56}$Fe is calculated by solving the gamma-ray transport with
   the approximation of an effective absorption opacity of
   0.06\,$Y_\mathrm{\!e}\,$cm$^2\,$g$^{-1}$, whereas positrons are assumed to
   deposit their energy locally.
The Compton electrons, occurring in scattering of gamma rays, lose their energy
   through Coulomb heating of free electrons, and ionization and excitation of
   atoms and ions.
The rates of the corresponding non-thermal processes are taken from
   \citet{KF_92}.

A non-equilibrium radiation field and a non-thermal excitation and ionization
   require solving the general problem of the level populations and the
   ionization balance instead of using the Boltzmann formulae and
   the Saha equations under the LTE conditions.
Multiple calculations of the corresponding equation of state in hydrodynamic
   modeling are possible by neglecting the excited atomic and ionic levels,
   i.e., by considering only the atomic and ionic ground states and their
   ionization balance.
The non-LTE ionization balance includes the elements H, He, C, N, O, Ne, Na,
   Mg, Si, S, Ar, Ca, Fe, and the negative hydrogen ion H$^{-}$.   It is
   controlled by the following elementary processes: photoionization and
   radiative recombination, electron ionization and three-particle
   recombination, and non-thermal ionization.
These processes are calculated using the partition functions of atoms and ions
   \citep{Irw_81}, the photoionization cross sections \citep{VY_95, VFKY_96},
   and the electron collisional ionization rates \citep{Vor_97}.
The photoionization of the negative hydrogen ion is treated with the cross
   section data of \citet{Wis_79}, and the electron collisional detachment
   reaction with the rate coefficient of \citet{JLE_87}.

The mean opacities, the thermal emission coefficient, and the contribution of
   lines to the opacity are computed taking into account non-LTE effects.
Photoionization, free-free absorption, Thomson scattering on free electrons,
   and Rayleigh scattering on neutral hydrogen contribute to the mean opacities.
For all atoms and ions but the negative hydrogen ion the free-free absorption
   coefficient is calculated with the effective nuclear charge corrected for
   screening effects \citep{SD_93} and the temperature-averaged free-free
   Gaunt factor \citep{Sut_98}.
For the negative hydrogen ion the free-free absorption coefficient was computed
   by \citet{BB_87}.
The calculation of the Rayleigh scattering by hydrogen atoms is based on the
   the cross section of \citet{Gav_67} and the exact static dipole
   polarizability of hydrogen from \citet{TP_71}.

In the outer, semitransparent and transparent layers, the ground state
   populations are calculated in non-LTE for the equation of state and for
   continuum opacity.
An expanding SN ejecta with a velocity gradient makes the contribution of
   spectral lines to the opacity essential and permits us to treat line
   opacities in the Sobolev approximation using the generalized formula
   of \citet{CAK_75}.
The expansion line opacities are determined by atomic and ionic level
   populations with the Boltzmann formulae and the Saha equations for
   a mixture of all elements from H to Zn at the local non-equilibrium
   radiation temperature.
The line database of \citet{Kur_02} provides us with oscillator strengths of 
   nearly $530,\!000$ lines.
The corresponding energy level data are from the atomic spectra database of
   the National Institute of Standards and Technology.

%% file: Results.tex
\section{Results}
\label{Results}
\subsection{Comparison between binary merger and single star pre-supernova models}

\begin{figure*}
    \centering
   \includegraphics[width=0.9\textwidth, clip, trim=20 379 47 208]{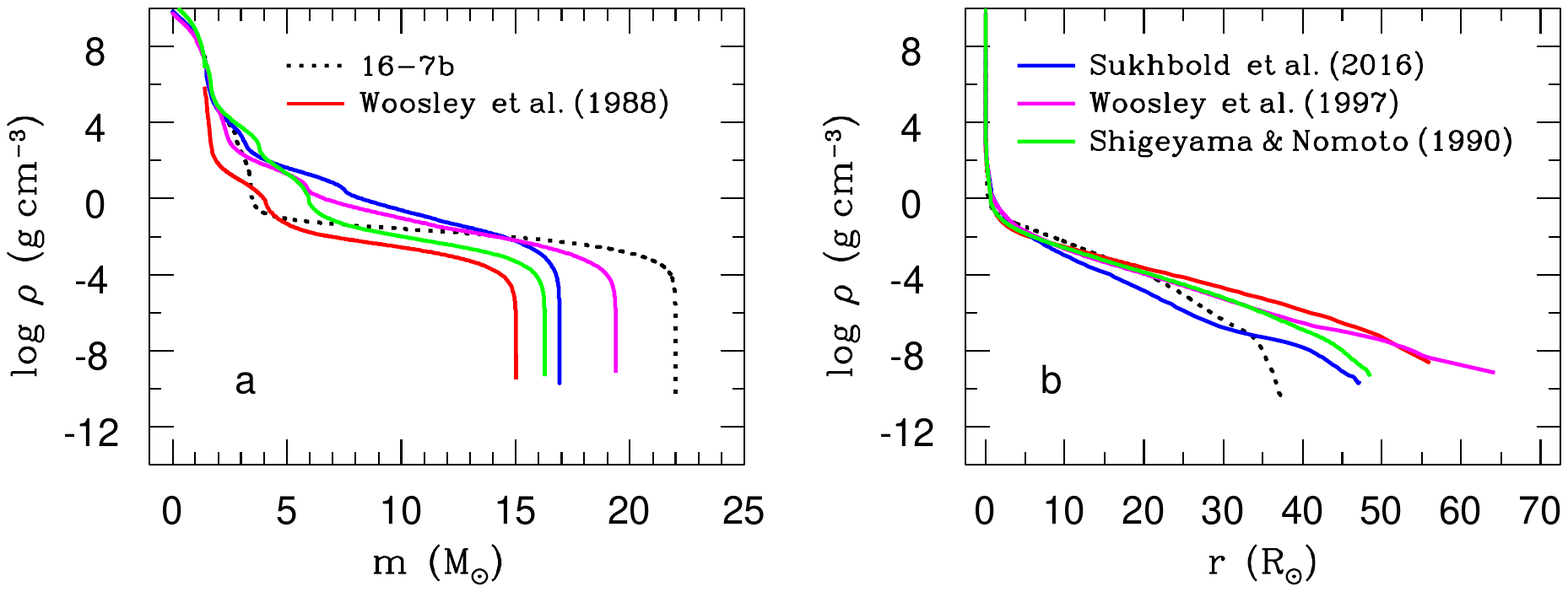}
    \caption{Density profiles of pre-SN models as a function of mass coordinate (\textsl{a}) and radius (\textsl{b}), including single star models from previous works (\textsl{solid lines}) and the merger model, Model \texttt{16-7b}, from the present work (\textsl{dotted black line}).}
    \label{den}
\end{figure*}

The binary merger BSG models (Table~\ref{presn2}) are structurally different
   from single star models.
In Fig.~\ref{den}, we compare the density profiles of the pre-SN Model \texttt{16-7b},
   which has a total mass of $22\,\Msun$, and single star pre-SN models with main-sequence masses of $15\,\Msun-20\,\Msun$ used in \citet{utrobin2015}. 

In general, our merger models have smaller He cores than those single star models ($3\,\Msun-4\,\Msun$ compared to $4\,\Msun-7.4\,\Msun$), more massive envelopes
   ($16\,\Msun-19.2\,\Msun$ compared to $9.5\,\Msun-13.6\,\Msun$) and have smaller radii as well
   ($29\,\Rsun-64\,\Rsun$ compared to $47\,\Rsun-64\Rsun$).
The merger models also have denser envelopes and a steeper density profile at the He-core/H-envelope interface than single star models (Fig.~\ref{den}). The compactness parameter \citep{ott2008,ott2013} in this study is calculated as  $\xi_{1.5}= (M/\Msun)/(R(M)/1000\,\mathrm{km})$ for $M=1.5\,\Msun$ as in \citet{utrobin2015}. Our models have a smaller core compactness parameter, between $0.006-0.011$,
   compared to single star models, which have a compactness parameter of $\xi_{1.5}=0.24-0.83$.
The iron-core masses of the merger pre-SN models are between $1.4\,\Msun-1.5\,\Msun$, whereas those of the single star models are between $1.2\,\Msun-1.5\,\Msun$.

The ``optimal'' non-evolutionary model computed by \citet{utrobin2005} whose
   explosion matched the light curve and photospheric velocity profile,
   had an ejecta mass of $18\,\Msun$ and a radius of $35\,\Rsun$.
The single star evolutionary models thus have larger radii and smaller ejecta
   masses than this optimal model, whereas some of our binary merger models
   from Table~\ref{presn2} do fit these ideal progenitor characteristics. Of these, the explosion of Model \texttt{16-7b}, with $R=37.4\,\Rsun$, $M_\mathrm{env}=18.8\,\Msun$, and $M_\mathrm{He,c}=3.4\,\Msun$ was found to best fit the light curve of SN~1987A. In the following sections, we first analyse how the light curve is affected by varying the explosion parameters for this model and after choosing the most optimal explosion parameter set, we show how the light curve responds to changing the properties of the progenitor.

\begin{figure*}
    \centering
    \includegraphics[width=0.9\textwidth, clip, trim=20 379 40 208]{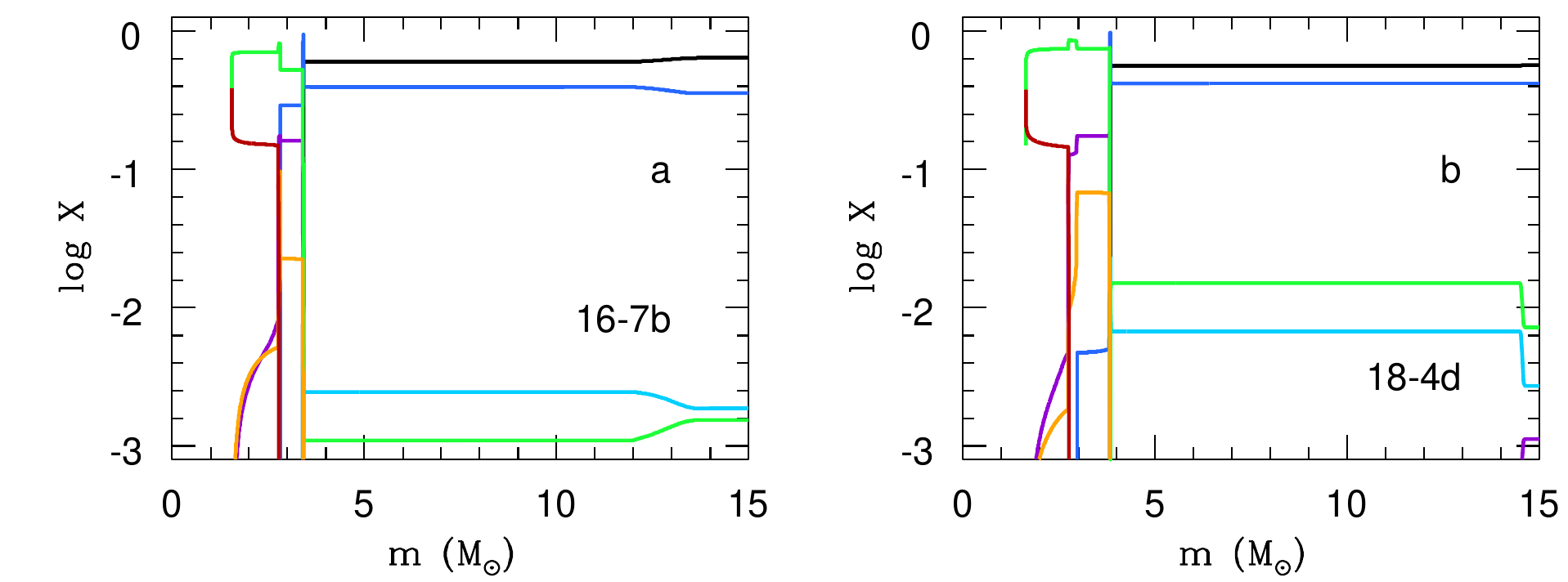}
   \hfill
    \caption{Chemical composition of the original pre-SN Models \texttt{16-7b} (a) and \texttt{18-4d} (b).  Mass fraction of hydrogen (\textsl{black line}), helium (\textsl{blue line}),  carbon (\textsl{violet line}), nitrogen (\textsl{cyan line}), oxygen (\textsl{green line}), neon (\textsl{orange line}), silicon (\textsl{firebrick line}), and iron (\textsl{red line}).}
    \label{presn-comp}
\end{figure*}

\subsection{The light curve of SN~1987A}

The light curves of Type IIP supernovae are powered by two sources:
   the deposited shock energy from the explosion and the energy from the
   radioactive decay of $^{56}\mathrm{Ni}$ to $^{56}\mathrm{Co}$ and later
   the decay of $^{56}\mathrm{Co}$ to $^{56}\mathrm{Fe}$.
In the case of compact progenitors, most of the shock energy is expended to
   adiabatically expand the star and hence the light curve is predominantly
   powered by gamma-ray energy \citep{woosley1988b,blinnikov2000, utrobin2004}.
If there was no $^{56}\mathrm{Ni}$ in the star, the light curve would begin
   to descend after $20\,\mathrm{days}-40\,$days as it radiates away the fraction of shock energy
   deposited \citep{woosley1988b,utrobin2004}. 

Subsequent to the triggering of the explosion, the shockwave moves out of the  centre towards the stellar surface, heating up the envelope as it does so, and causes it to expand with velocities increasing outward and exceeding the local escape velocity.
After the shock breaks out of the surface, the luminosity surges to a peak between $10^{43.5}-10^{45}\,$erg$\,$s$^{-1}$ \citep{woosley1988b,blinnikov2000,utrobin2004}. As the star expands and cools, the luminosity decreases until about    $\sim8\,$days at which point the temperature inside the star reaches that of hydrogen recombination. A cooling and recombination wave (CRW) forms and travels inward, with the photosphere in it, and the internal energy of the star begins to radiate outward \citep{utrobin1993}. The period of the first $\sim30\,$days is referred to as the \textsl{early light curve}. From Day $\sim8$ to Day $\sim30$, the SN luminosity is determined by the CRW
   properties. After about $30\,$days and until about $120\,$days, referred to as the \textsl{middle light curve},
   the energy diffused outward increasingly becomes dominated by the gamma-ray
   radioactive energy \citep{utrobin1993}.
When the SN ejecta become optically thin for radiation, the bolometric luminosity
   follows the instantaneous release of energy from the decay of
   $^{56}\mathrm{Co}$ to $^{56}\mathrm{Fe}$.
This is the \textsl{late light curve} phase ($t>120\,$days). 

\subsubsection{Varying explosion parameters}

In this section, we study the effect of varying the explosion energy, the width of boxcar
   mixing, and the velocity of nickel mixed to the surface on the light curve
   shape. For the explosion of all progenitor models, we fix the nickel mass at $0.073\,\Msun$ and the mass cut at $1.4\,\Msun$.

\begin{figure*}
\centering
   \includegraphics[width=0.9\textwidth, clip, trim=20 379 47 208]{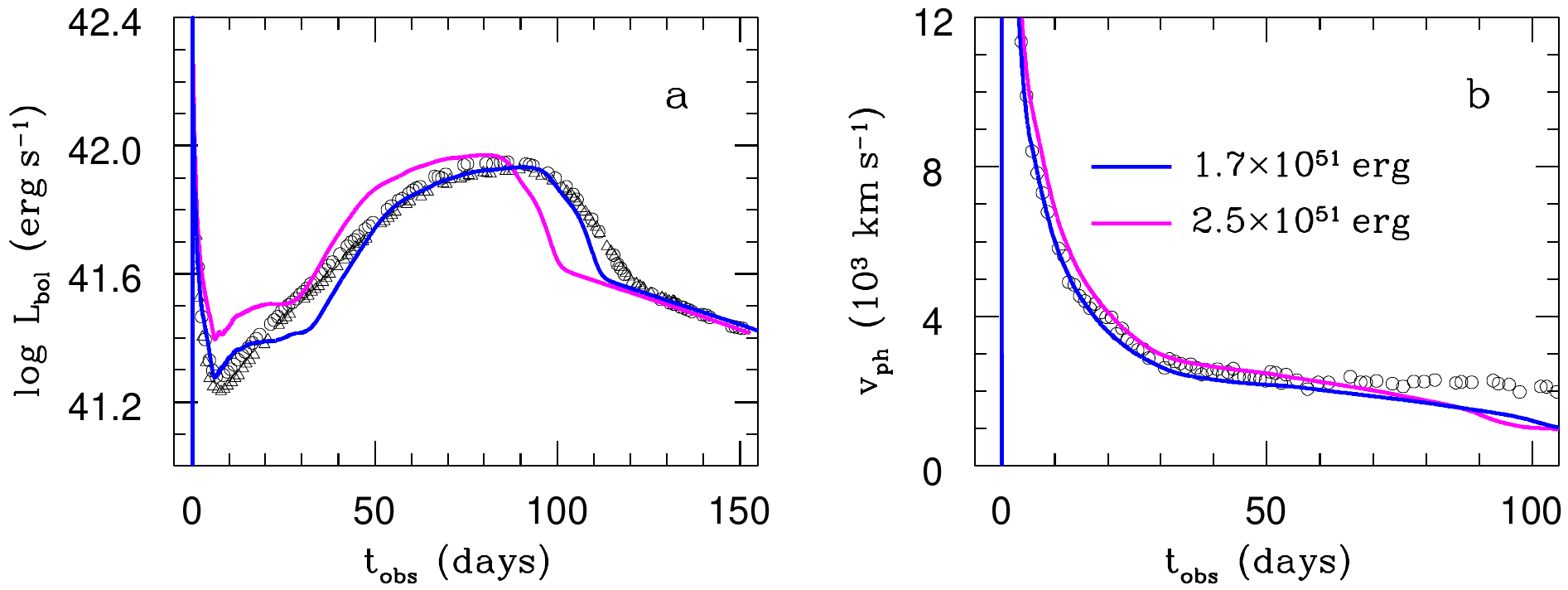}
   \caption{
   Dependence on the explosion energy.
   Bolometric light curves (\textsl{a}) and photospheric velocity (\textsl{b}) as a function
      of time of Model \texttt{16-7b} for the explosion energies of
      $1.7\times10^{51}\,$erg (\textsl{blue line}) and $2.5\times10^{51}\,$erg
      (\textsl{magenta line}), respectively.
   The computed curves are compared with the observed bolometric luminosity of
      SN~1987A obtained by \citet[][\textsl{open circles}]{CMM_87, CWF_88} and
      \citet[][\textsl{open triangles}]{HSGM_88}, and with the velocity at the
      photosphere estimated by \citet{PHHN_88} using the absorption minimum of
      the $\ion{Fe}{ii}\sim$5169\,\AA\ line (\textsl{open circles}).}
   \label{fig:var_E}
\end{figure*}

\begin{figure*}
\centering
   \includegraphics[width=0.9\textwidth, clip, trim=36 380 45 208]{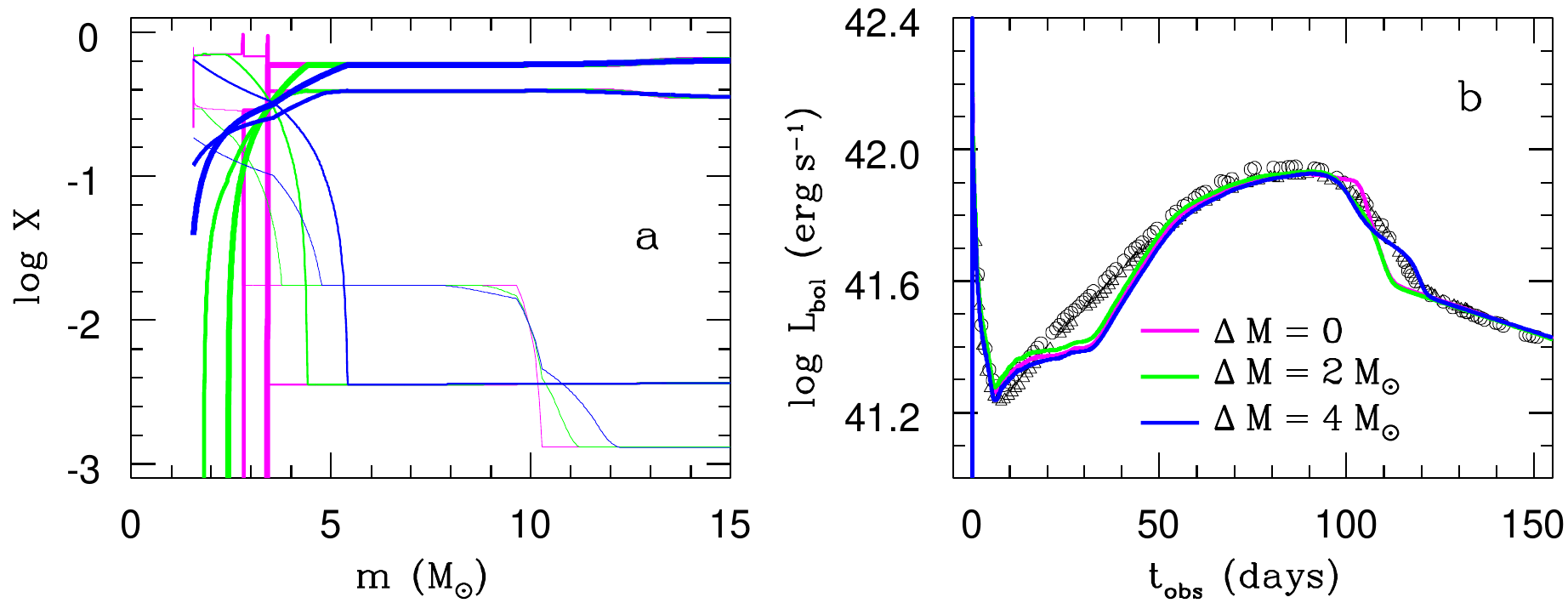}
   \caption{
   Dependence on the boxcar mass width $\Delta{}M$.  \textsl{Panel a} shows the chemical composition of the pre-SN models based on Model \texttt{16-7b} after the boxcar averaging with the mass width $\Delta{}M=2\,\Msun$       (\textsl{green lines}) and $4\,\Msun$ (\textsl{blue lines})  and \textsl{Panel b} shows the corresponding bolometric light curves compared with the observations of SN~1987A       obtained by \citet[][\textsl{open circles}]{CMM_87, CWF_88} and \citet[][\textsl{open triangles}]{HSGM_88}.  For reference, Model \texttt{16-7b} without the boxcar averaging of the chemical      composition is shown with $\Delta{}M=0$ (\textsl{magenta lines}).  Mass fraction of hydrogen (\textsl{thick line}), helium (\textsl{medium line}), CNO elements (\textsl{thin line}), and Fe-peak elements (\textsl{tiny line}).
   }
   \label{fig:var_dM}
\end{figure*}

\begin{figure*}
\centering
   \includegraphics[width=0.9\textwidth, clip, trim=36 380 45 208]{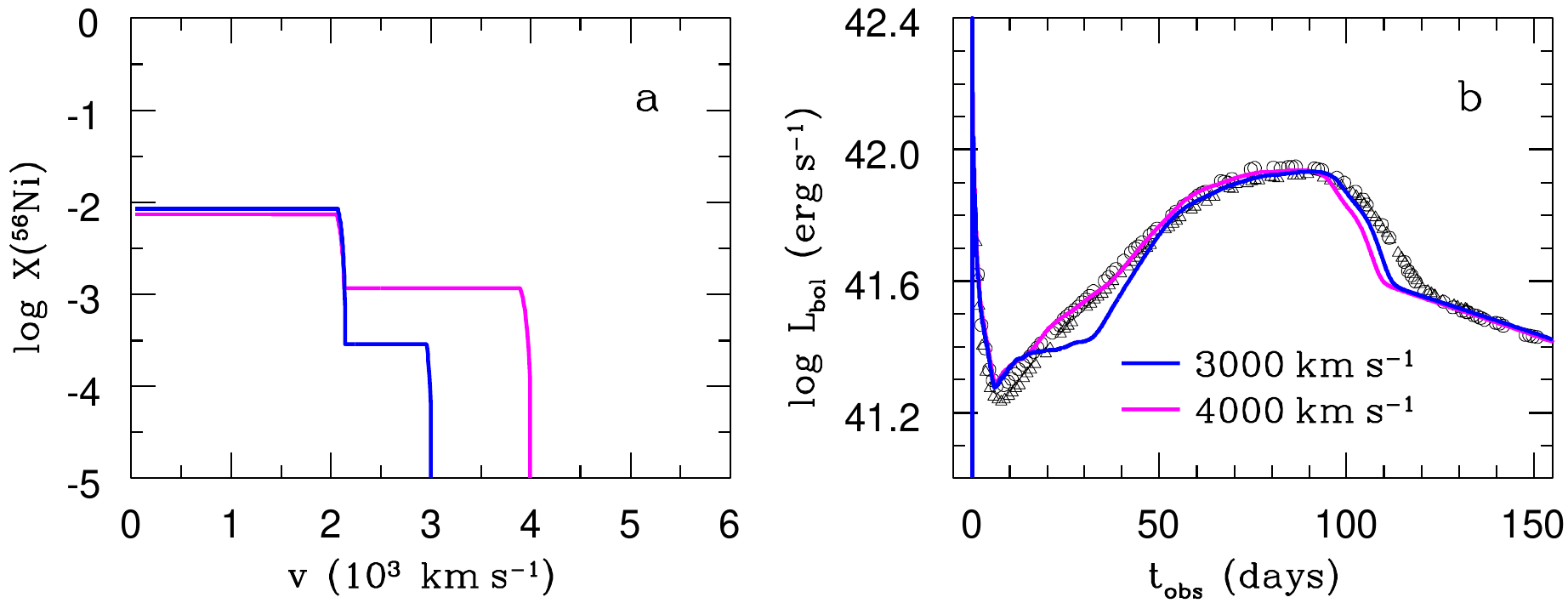}
   \caption{
   Dependence on the extent of mixing of radioactive $^{56}$Ni.
   Mass fraction of radioactive $^{56}$Ni as a function of velocity at Day 50
      for  Model \texttt{16-7b} (\textsl{a}) and the corresponding bolometric light curves (\textsl{b})
      compared with the observations of SN~1987A obtained by
      \citet[][\textsl{open circles}]{CMM_87, CWF_88} and \citet[][\textsl{open triangles}]{HSGM_88}.  
   }
   \label{fig:var_Vni}
\end{figure*}

All explosion parameters in this section are varied for Model \texttt{16-7b} of
   Table~\ref{presn2}.
In Fig.~\ref{fig:var_E}, we vary the explosion energy between $1.7\,\B-2.5\,\B$.
The best match for the light curve is produced for $E=1.7\,\B$, at the dip at
   $\sim8\,$days and the overall fit in the middle light curve region.
This result is in concurrence with the prediction of \citet{utrobin2005} for
   their optimal model, in which the $\mathrm{H}_\mathrm{\alpha}$ line profile
   at $4.64\,$days reproduced the spectroscopic observations of SN~1987A at
   a ratio of $E/M_\mathrm{ej}=1.5\,$B/$18\,\Msun$.
We hence use the same $E/M_\mathrm{ej}$ value for all our hydrodynamic models
   in this study.  

Next, we use an artificial procedure that mimics the mixing that arises from instabilities in the ejecta as seen in 3D simulations after the passage of the shockwave \citet{kifonidis2003,kifonidis2006,wong2015} 
and which has also been used in other 1D explosion studies such as those of \citet{pinto1988,kasen2009,dessart2013,morozova2015}.  This 1D mixing is implemented via a running boxcar of width $\Delta{}M$ through the envelope of the star, which averages the composition of all chemical species other than $^{56}\mathrm{Ni}$.  In Fig.~\ref{fig:var_dM}, we vary the width of the running boxcar, $\Delta{}M$, which effectively smoothens composition gradients in the ejecta.  Increasing $\Delta{}M$ increases the spread of hydrogen inward (\textsl{thick lines} in Fig.~\ref{fig:var_dM}).
The boxcar width determines the distribution of hydrogen within the He core,
   which, in turn, determines the shape of the dome at the transition from the
   luminosity maximum to the radioactive tail (Fig.~\ref{fig:var_dM}b). 
A value of $\Delta{}M=2\,\Msun$ mimics hydrogen mixing down to zero velocity,
   which agrees with spectral observations, and produces the best match with
   the dome shape of SN~1987A.

After setting the values of $E$ and $\Delta{}M$, we selectively vary the mixing of $^{56}\mathrm{Ni}$ in velocity space, keeping all other species intact (Fig.~\ref{fig:var_Vni}).  We find that in order to have a smoothly rising light curve to maximum, a strong nickel mixing velocity of $4,\!000\,\kms$ is required, which is
   the same result as found in previous explosion studies using evolutionary
   models \citep{shigeyama1990,blinnikov2000,utrobin2004}.  Using the observed value of $3,\!000\,\kms$ results in a slower luminosity increase between
   $15\,\mathrm{days}-35\,$days due to the photosphere approaching the region where
   the internal energy of the ejecta has been radiated away and the gamma-ray
   energy from the radioactive decay has not been diffused yet.
If $^{56}\mathrm{Ni}$ is distributed out to regions moving at larger
   velocities, then the gamma rays diffused compensates for the drop in
   internal energy of the envelope and the light curve rises smoothly.

\subsubsection{Varying progenitor parameters}

From the results of the previous section, we choose the following explosion
   parameters for the investigation into other progenitor models:
   $E/M_\mathrm{ej}=1.5\,$B/$18\,\Msun$, nickel mixing velocity of $3,\!000\,\kms$
   and boxcar mixing of $\Delta{}M=2\,\Msun$. 

\begin{figure*}
\centering
   \includegraphics[width=0.9\textwidth, clip, trim=29 162 45 211]{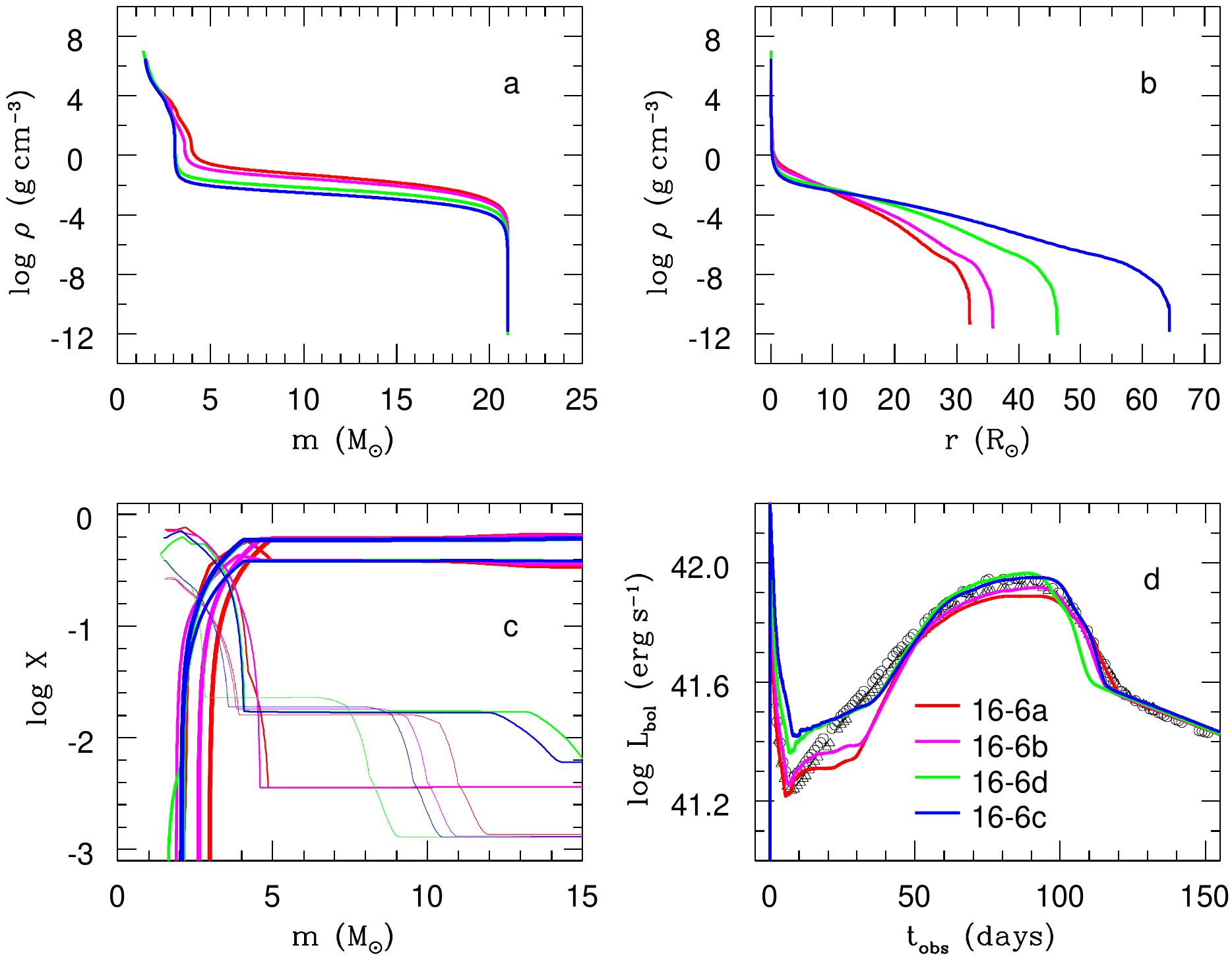}
   \caption{
   Variation in the fraction of the He shell of the He core dredged up,
      $f_\mathrm{sh}$: 10\,\% (\textsl{red lines}), 50\,\% (\textsl{magenta lines}),
      90\,\% (\textsl{green lines}), and 100\,\% (\textsl{blue lines}) for $M_1=16\,\Msun$
      and $M_2=6\,\Msun$.
   \textsl{Panel a}: Density distributions as functions of interior mass for Models
      \texttt{16-6a}, \texttt{16-6b}, \texttt{16-6d}, and \texttt{16-6c}.
   \textsl{Panel b}: Density distributions as functions of radius for the same models.
   \textsl{Panel c}: Chemical composition of the corresponding pre-SN models after
      the boxcar averaging with the mass width $\Delta{}M=2\,\Msun$.
   Mass fraction of hydrogen (\textsl{thick line}), helium (\textsl{medium line}), CNO elements
      (\textsl{thin line}), and Fe-peak elements (\textsl{tiny line}).
   \textsl{Panel d}:  Calculated bolometric light curves are overplotted
      on the bolometric data of SN~1987A obtained by \citet[][\textsl{open circles}]{CMM_87, CWF_88}
       and \citet[][\textsl{open triangles}]{HSGM_88}.
   }
   \label{fig:16-6}
\end{figure*}

\begin{figure*}
\centering
   \includegraphics[width=0.9\textwidth, clip, trim=29 162 45 211]{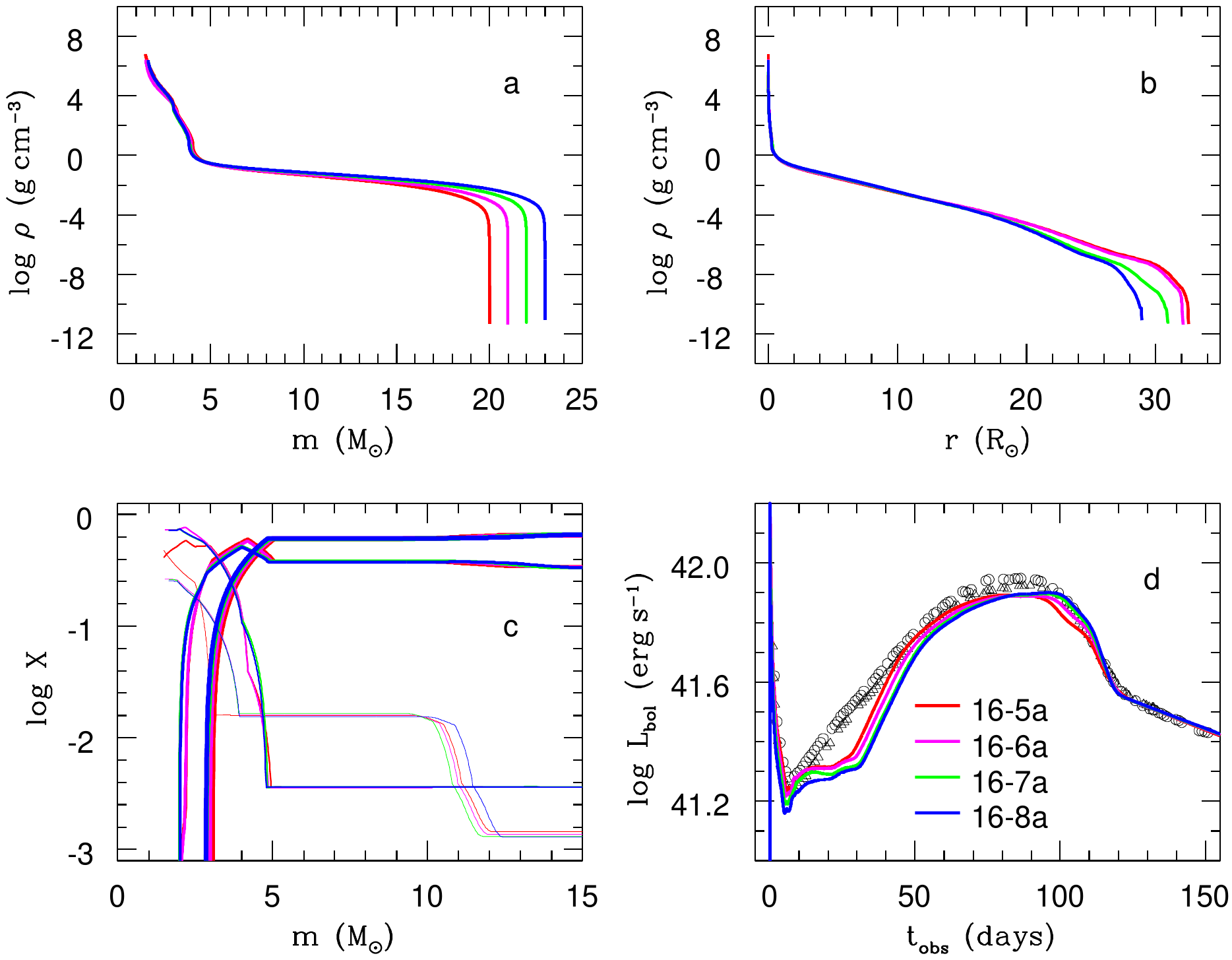}
   \caption{
   Variation in the secondary mass: $M_2=5\,\Msun$ (\textsl{red lines}),
      $6\,\Msun$ (\textsl{magenta lines}), $7\,\Msun$ (\textsl{green lines}), and
      $8\,\Msun$ (\textsl{blue lines}) for the primary mass $M_1=16\,\Msun$
      and the fraction $f_\mathrm{sh}=10\,\%$.
   See Figure~\ref{fig:16-6} legend for details.}
   \label{fig:16-a}
\end{figure*}

In Fig.~\ref{fig:16-6}, we show how the fraction of the He shell of
   the He core dredged up affects the light curve shape for merger models
   with a primary of mass $M_{1}=16\,\Msun$ and a secondary of mass
   $M_{2}=6\,\Msun$.
These are  Models \texttt{16-6a}, \texttt{16-6b}, \texttt{16-6d}, and \texttt{16-6c} in Table~\ref{presn2}, which have the same pre-SN mass and have $f_\mathrm{sh}$ set as $10\,\%$, $50\,\%$, $90\,\%$, and $100\,\%$, respectively. With increasing $f_\mathrm{sh}$, the models have larger radii (Fig.~\ref{fig:16-6}\textsl{b}).  In addition, in this sequence of the models, the He-core mass decreases and,
   consequently, the location of the H-rich envelope is more inward, whereas
   in the outer ($m>10\Msun$) H-rich layers the hydrogen abundance decreases
   and the helium abundance increases (Fig.~\ref{fig:16-6}\textsl{c}).
The greater the pre-SN radius is, the longer is the expansion timescale for
   the envelope.
Hence the CRW forms sooner for smaller initial radii and the light curve begins
   to ascend earlier.
More important is that increasing the initial radius reduces the role of
   adiabatic energy losses \citep{utrobin2005}, resulting in a higher internal
   energy per unit mass, with roughly the same internal energy per unit mass
   initially deposited by the shock, and consequently a higher luminosity
   between $10-35\,$days that causes the luminosity to increase with a less
   pronounced bump (Fig.~\ref{fig:16-6}\textsl{d}). 
Thus, overall, with increasing $f_\mathrm{sh}$, the luminosity rises earlier and shifts upward during the early light curve phase.  

In Fig.~\ref{fig:16-a}, we show how the accreted secondary mass $M_{2}$,
   affects the light curve shape. Shown are  Models \texttt{16-5a}, \texttt{16-6a}, \texttt{16-7a}, and \texttt{16-8a} of Table~\ref{presn2}, which
   have $f_\mathrm{sh}=10\,\%$ and $M_{2}=5\,\Msun$, $6\,\Msun$, $7\,\Msun$, and $8\,\Msun$, respectively, and radii in the range $28.8\,\Rsun-32.8\,\Rsun$.
The density profiles of the cores of these models are nearly the same (Fig.~\ref{fig:16-a}\textsl{a}), as also are their He-core masses (Table~\ref{presn2}). Increasing $M_{2}$ increases the ejecta mass, but shrinks the pre-SN radius.
The latter increases the role of adiabatic energy losses during the supernova explosion, shifting the luminosity downward during the early light curve phase and delaying the rise of the light curve to its maximum because of a greater total internal energy, which is proportional to the ejecta mass, and this energy is then radiated away over a longer period of time. Thus overall, the light curve shifts to later times with increasing $M_{2}$,  but keeping nearly the same characteristic width of the dome. The behaviour of the bump feature between $10-35\,$days with increasing $M_{2}$ (Fig.~\ref{fig:16-a}\textsl{d}) is similar to that with decreasing  (Fig.~\ref{fig:16-6}\textsl{d}).
In both cases, i.e., changing $f_\mathrm{sh}$ and $M_{2}$, the progenitor property that most affects the shape of the light curve is its pre-SN radius.

Of all progenitor models studied in this work, the one that best fits the observed properties of \sk and whose explosion matches the light curve and the photospheric velocity of iron absorption lines (Fig.~\ref{fig:SN1987A}) is Model \texttt{16-7b}, with an explosion energy of $1.7\,$B, ejecta mass of $20.6\,\Msun$, nickel mixing velocity of $3,\!000\,\kms$, and a nickel mass of $0.073\,\Msun$.

\subsection{Other peculiar Type IIP supernovae: SN~1998A and SN~2006V} 

Here we discuss the application of our analysis to two other peculiar Type IIP SNe. Table~\ref{predicted} lists the predicted progenitor properties from hydrodynamic models, semi-analytic models, and scaling relations for five peculiar Type IIP SNe. 

\begin{table*}
\caption{Predicted explosion and progenitor properties for five peculiar
      Type IIP supernovae. 
   $M_\mathrm{ej}$: ejecta mass, $R$: radius of progenitor, $E$: explosion
      energy, $M_\mathrm{Ni}$: nickel mass.
   In parenthesis are the values of SN~1987A against which these parameters
      have been calibrated.
   Nickel masses are taken from Tables 4 and 5 in \citet{lusk2017}.
   The final column lists the method used for the calculations.}
 \label{predicted}
 \begin{tabular}{llccccl}
  \hline
Supernova & Source & $M_\mathrm{ej}\,(\Msun)$ & $R\,(\Rsun)$ & $E\,(\B)$ & $M_\mathrm{Ni}\,(\Msun)$ & Method \\
  \hline
SN~1998A & \citet{pastorello2005} & 22 (18) & 86.3 (43.2) & 5-6 (1.6) & 0.11 (0.075) & Semi-analytic model\\
	      & \citet{pastorello2012} & & &  & 0.09 (0.075)& Updated results \\
SN~2000cb & \citet{kleiser2011} & 16.5 (14) & 43.2 (43.2) &  4 (1.1) & 0.1 & Scaling relations \\
	      &       & 17.5 (14) & 43.2 (43.2) &  2 (1) & $0.1\pm0.02$ & Hydrodynamic model\\
	      & \citet{utrobin2011} & 22.3 & $35\pm14$ & 4.4 & 0.083 & Hydrodynamic model \\
SN~2006V & \citet{taddia2012} & 20 (14) & $<50$ (33) &  & $<0.127$ (0.078) & Scaling relations\\
 	      &			& 17 (14) & 75 (33) & 2.4 (1.1) & 0.127 (0.078) & Semi-analytic model\\
SN~2006au & \citet{taddia2012} & 20 (14) & $<50$ (33)&  & $<0.127$ (0.078) & Scaling relations\\
 	      &			& 19.3 (14) & 90 (33) & 3.2 (1.1) & 0.073 (0.078) & Semi-analytic model\\
SN~2009E & \citet{pastorello2012} & 26 & 86.2 & 1.3 & 0.039 & Semi-analytic model \\
 	      &			&  19 & 100.7 & 0.6 & 0.043 & Hydrodynamic model\\
  \hline
 \end{tabular}
\end{table*}

\begin{figure*}
\centering
   \includegraphics[width=0.8\textwidth, clip, trim=29 162 45 211]{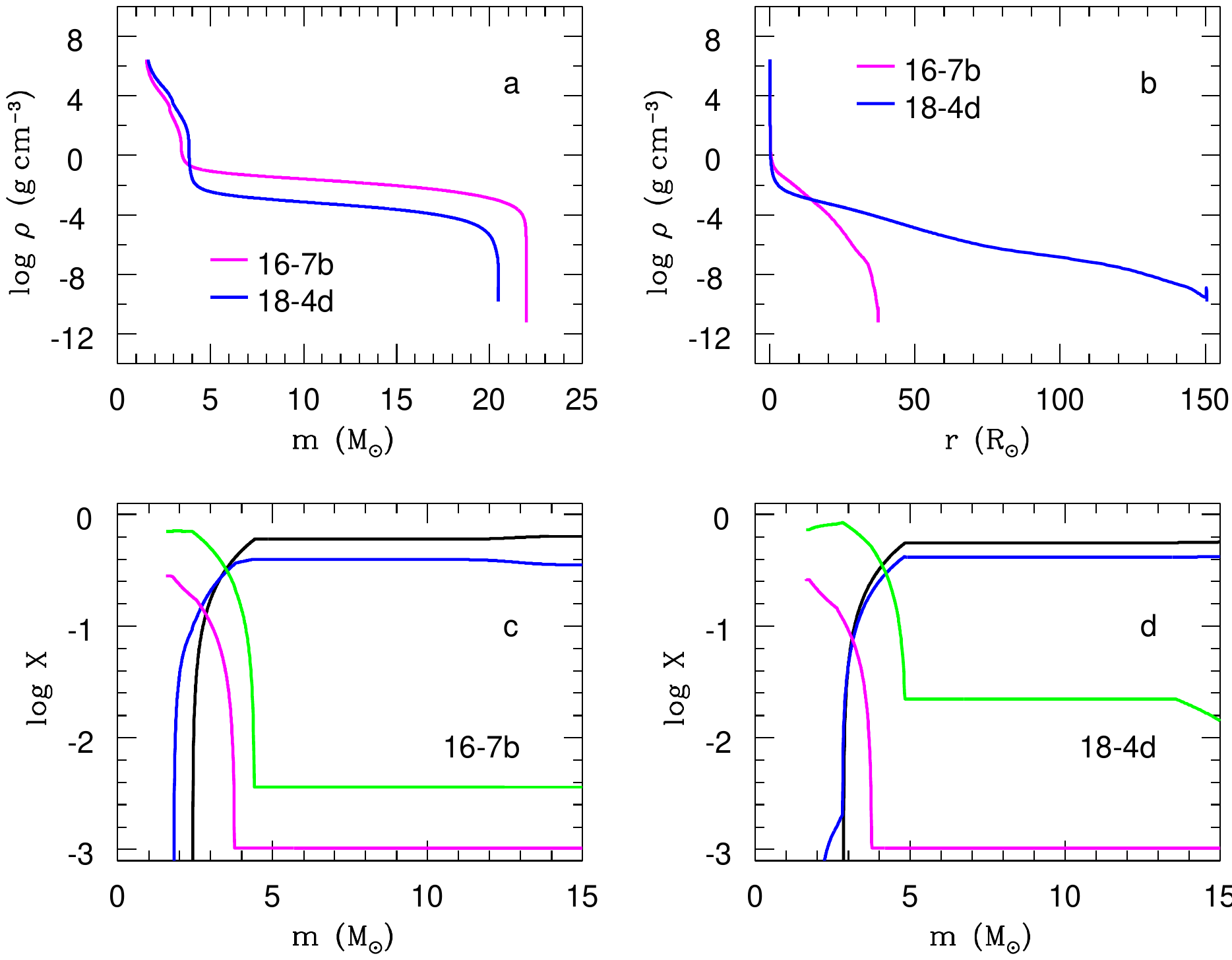}
   \caption{
   Pre-SN models for peculiar Type IIP supernovae.
   \textsl{Panel a}: Density distributions as functions of interior mass for Models
      \texttt{16-7b} (\textsl{magenta line}) and 18-4d (blue line).
   \textsl{Panel b}: Density distributions as functions of radius for the same models.
   \textsl{Panel c}: Chemical composition of the pre-SN Model \texttt{16-7b} after
      the boxcar averaging with the mass width $\Delta\,M=2\,\Msun$.
   Mass fraction of hydrogen (\textsl{black line}), helium (\textsl{blue line}), CNO elements
      (\textsl{green line}), and Fe-peak elements (\textsl{magenta line}).
   \textsl{Panel d}: The same for the pre-SN Model \texttt{18-4d}.
   }
   \label{fig:pSN-16-18}
\end{figure*}

\begin{figure*}
\centering
   \includegraphics[width=0.67\columnwidth, clip, trim=18 182 37 93]{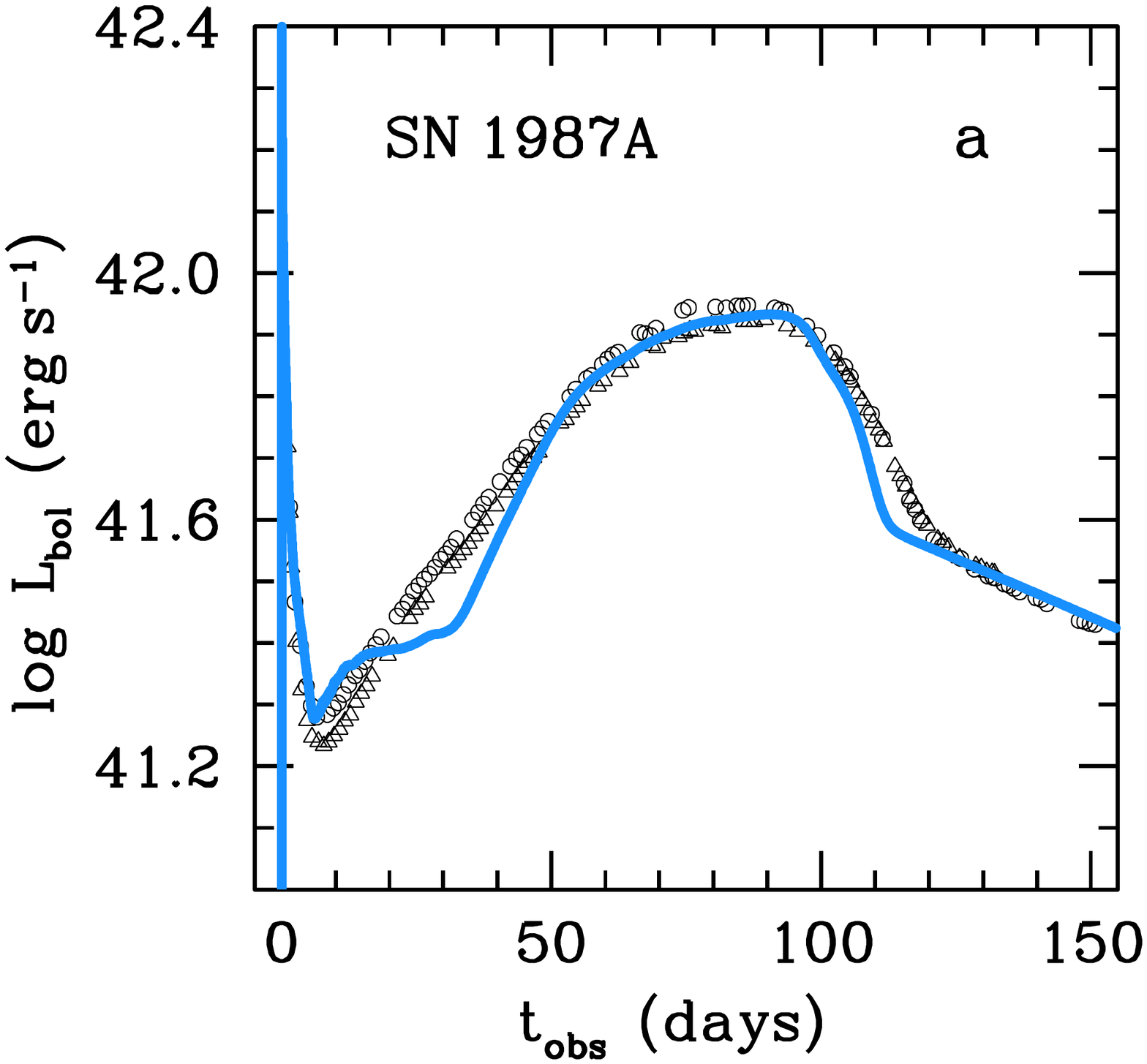}
   \includegraphics[width=0.67\columnwidth, clip, trim=18 182 37 93]{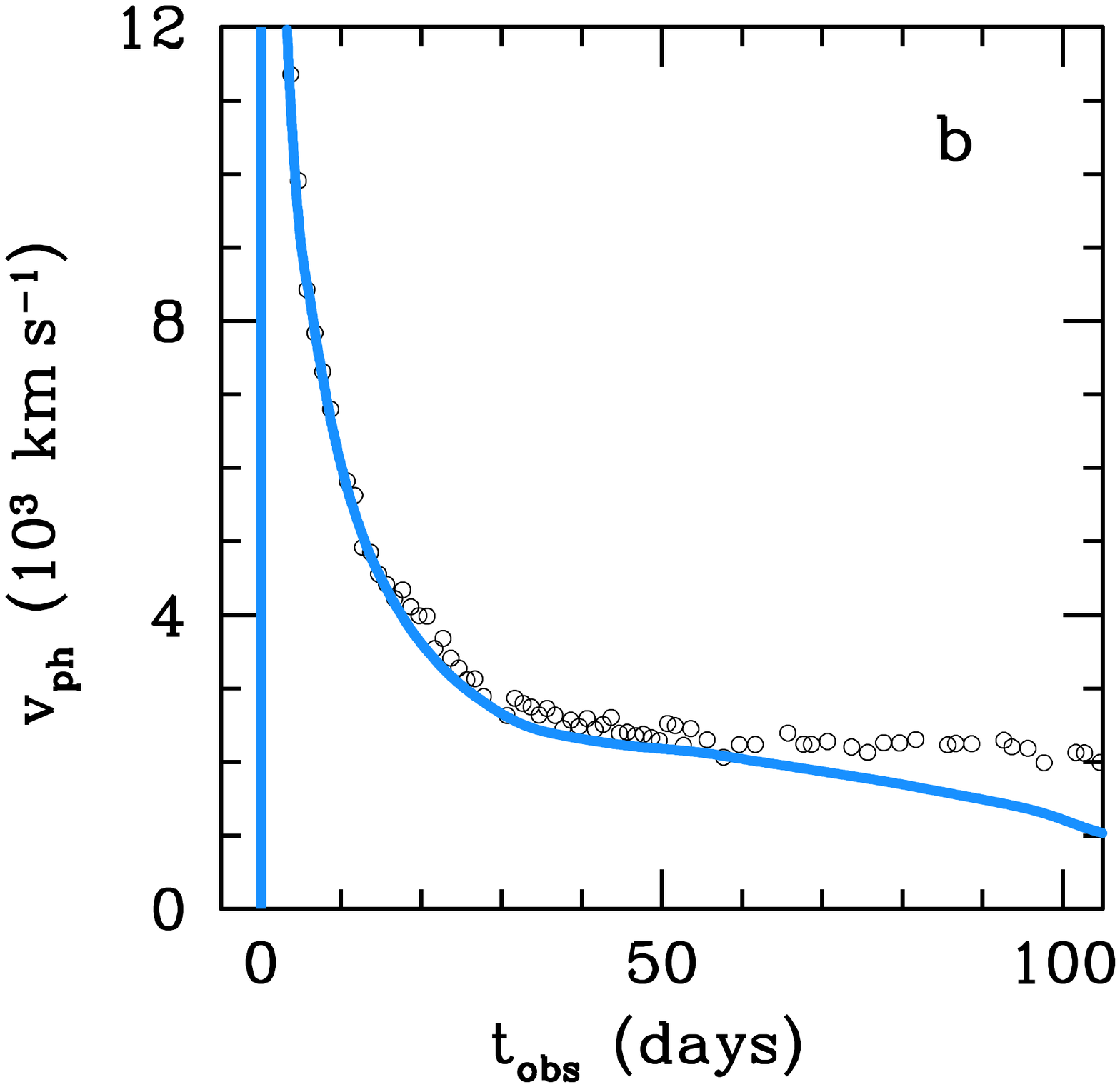}
   \includegraphics[width=0.67\columnwidth, clip, trim=18 182 37 93]{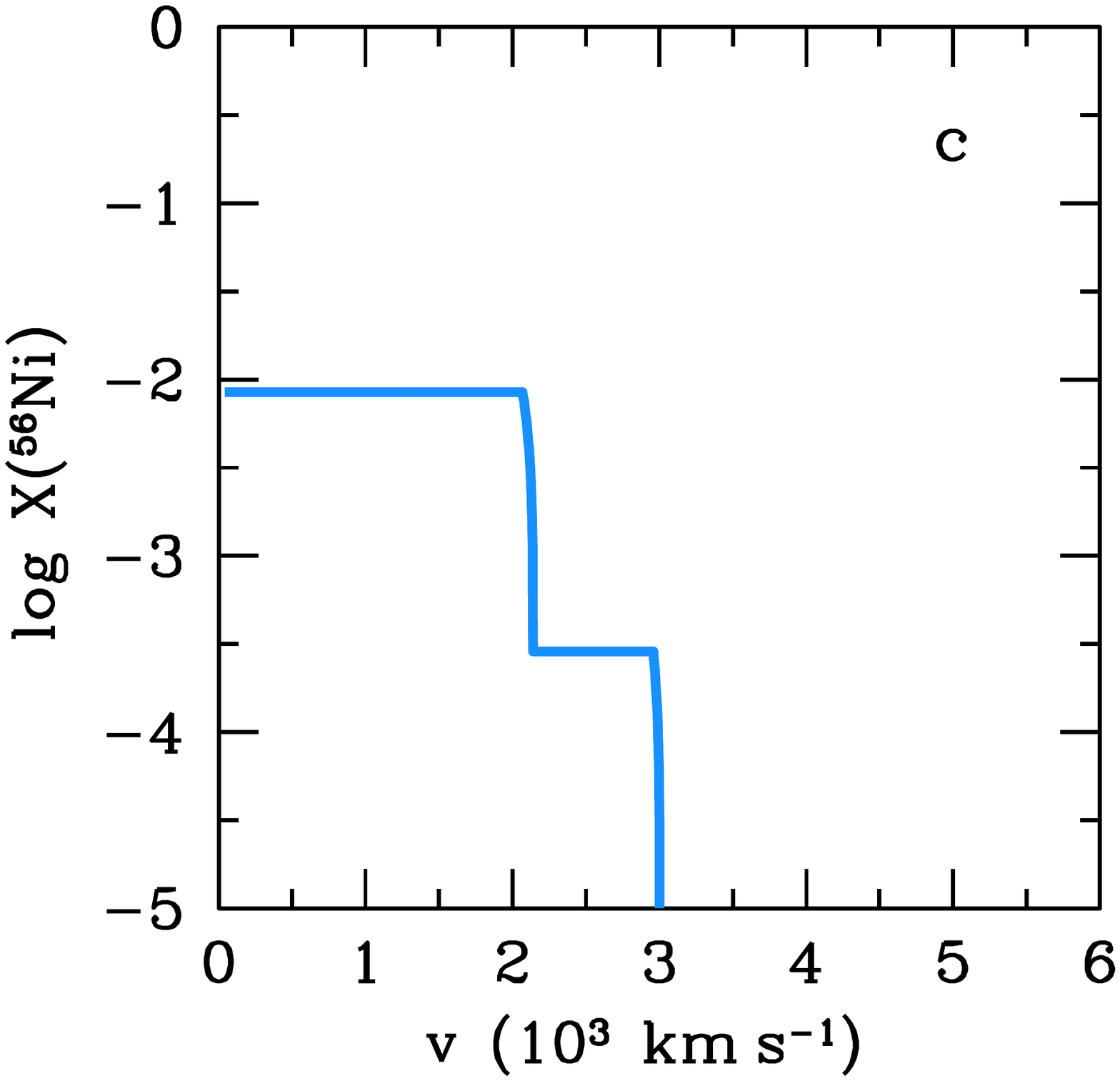}\\
   \caption{
   Hydrodynamic model for SN 1987A.
   \textsl{Panel a}:  Calculated bolometric light curve is compared with the
      observations of SN~1987A obtained by \citet[][\textsl{open circles}]{CMM_87, CWF_88}
      and \citet[][\textsl{open triangles}]{HSGM_88}.
   \textsl{Panel b}: Calculated photospheric velocity is overplotted on the  
      velocity at the photosphere estimated by \citet{PHHN_88} with the
      absorption minimum of the $\ion{Fe}{ii}~5169\,\AA$ line (\textsl{open circles}).
   \textsl{Panel c}: Mass fraction of radioactive $^{56}\mathrm{Ni}$ as a function of velocity at Day 50.
   }
   \label{fig:SN1987A}
\end{figure*}

\begin{figure*}
\centering
   \includegraphics[width=0.67\columnwidth, clip, trim=18 182 37 93]{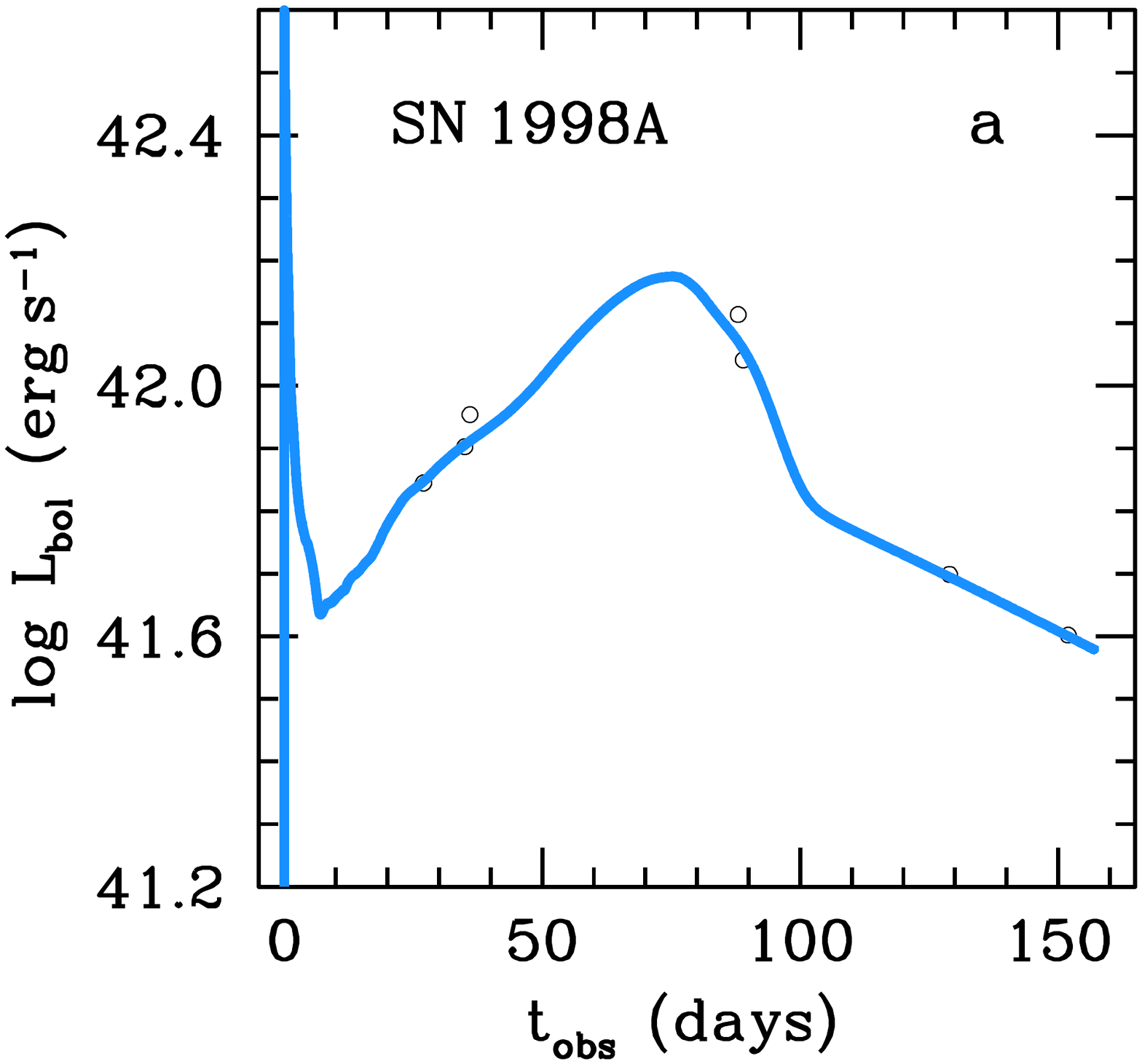}
   \includegraphics[width=0.67\columnwidth, clip, trim=18 182 37 93]{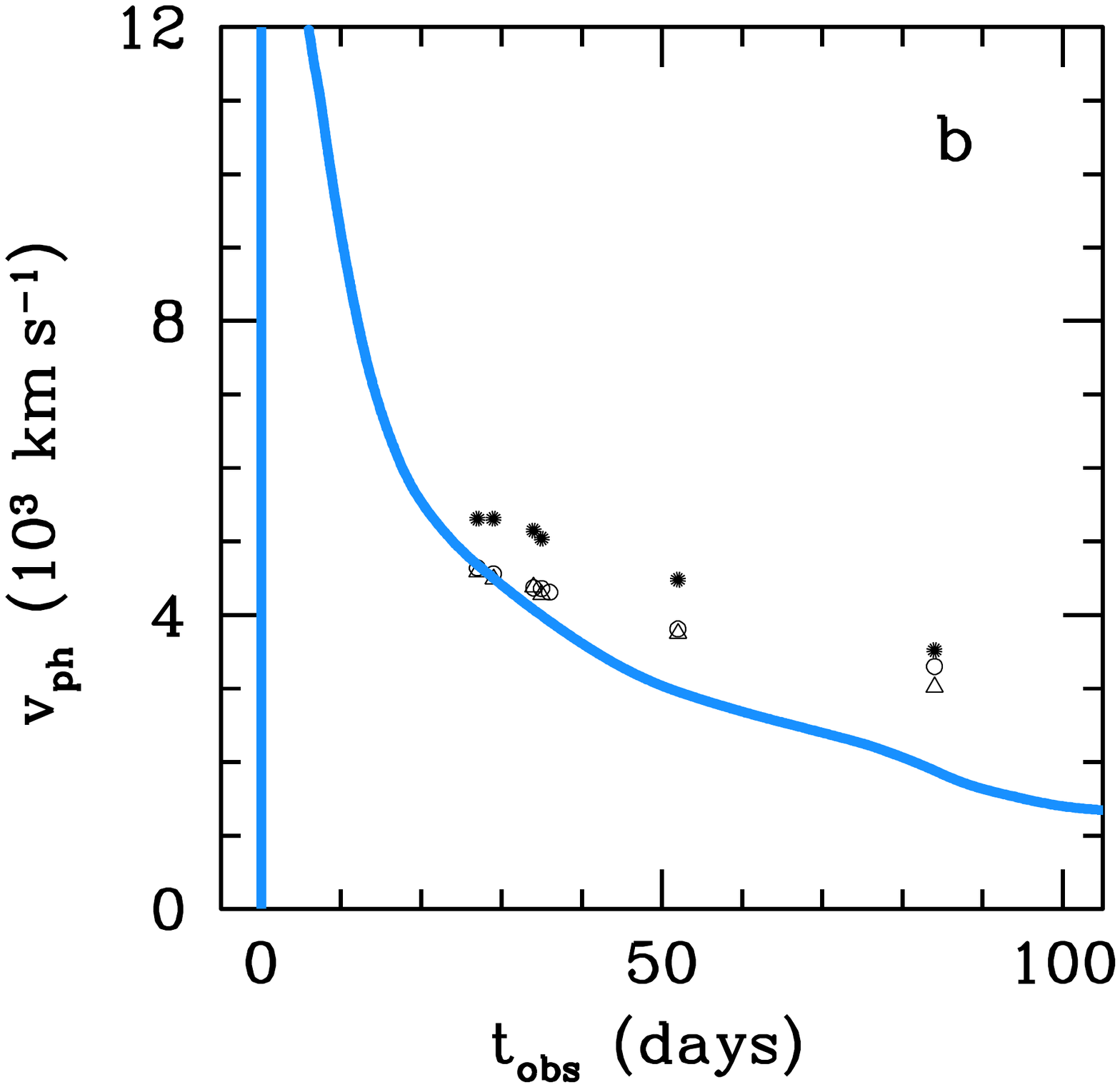}
   \includegraphics[width=0.67\columnwidth, clip, trim=18 182 37 93]{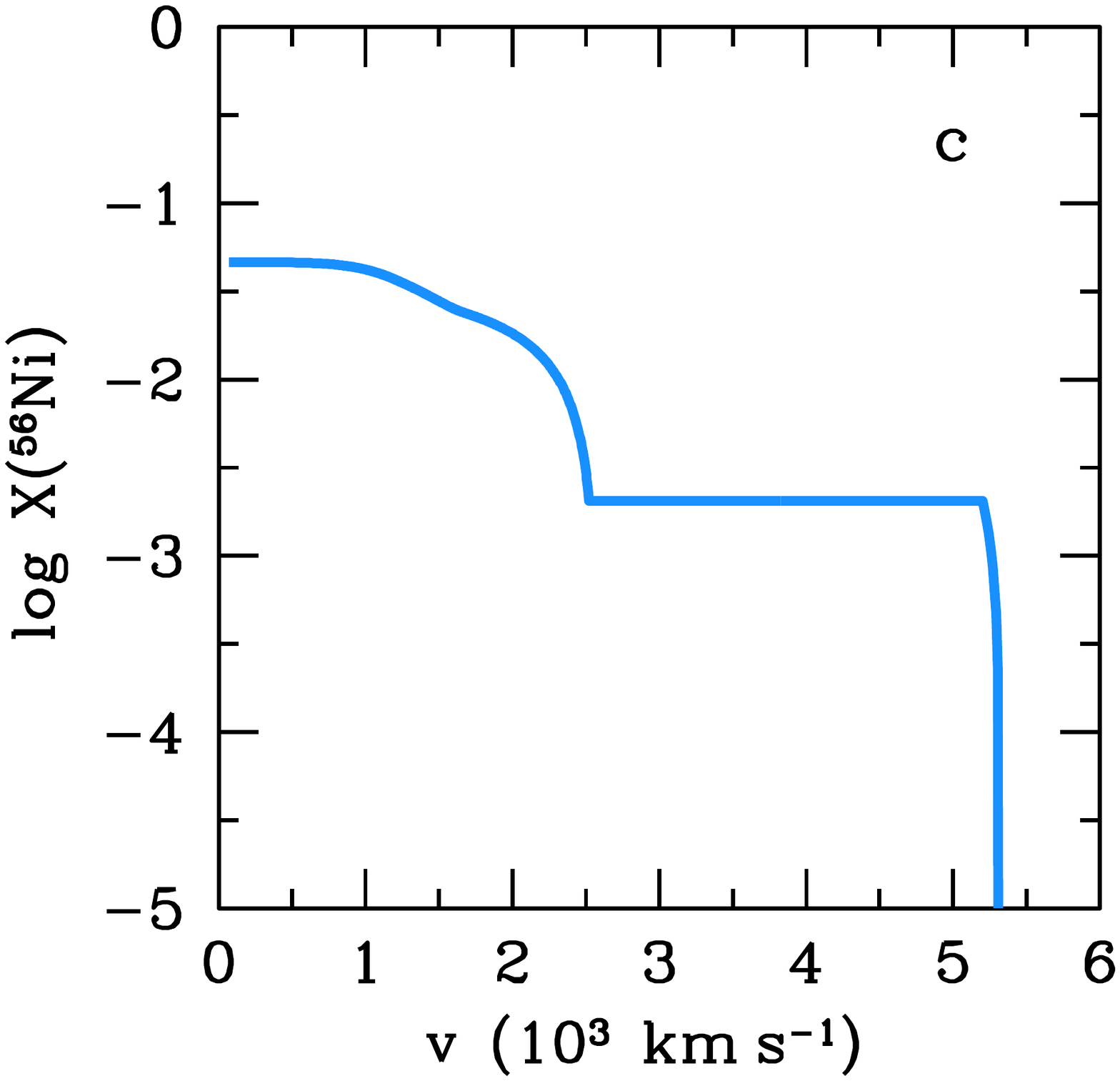}\\
   \caption{
   Hydrodynamic model for SN 1998A.
   \textsl{Panel a}: Calculated light curve is compared with the bolometric data
      of SN~1998A estimated by \citet{lusk2017} by means of direct integration
      (\textsl{open circles}).
   To fit the observations, the explosion date is suggested to be later by
      $10\,$days relative to that accepted by \citet{pastorello2005}.
   \textsl{Panel b}: Calculated photospheric velocity is overplotted on the  
      velocity at the photosphere estimated by \citet{pastorello2005} with the
      absorption minima of the $\ion{Ba}{ii}~6142\,\AA$ (\textsl{open circles}),
      $\ion{Fe}{ii}~5169\,\AA$ (\textsl{filled circles}), $\ion{Sc}{ii}~5527\,\AA$
     (\textsl{open triangles}) lines.
   \textsl{Panel c}: Mass fraction of radioactive $^{56}$Ni as a function of velocity at Day 50.
   }
   \label{fig:SN1998A}
\end{figure*}

\begin{figure*}
\centering
   \includegraphics[width=0.67\columnwidth, clip, trim=18 182 37 93]{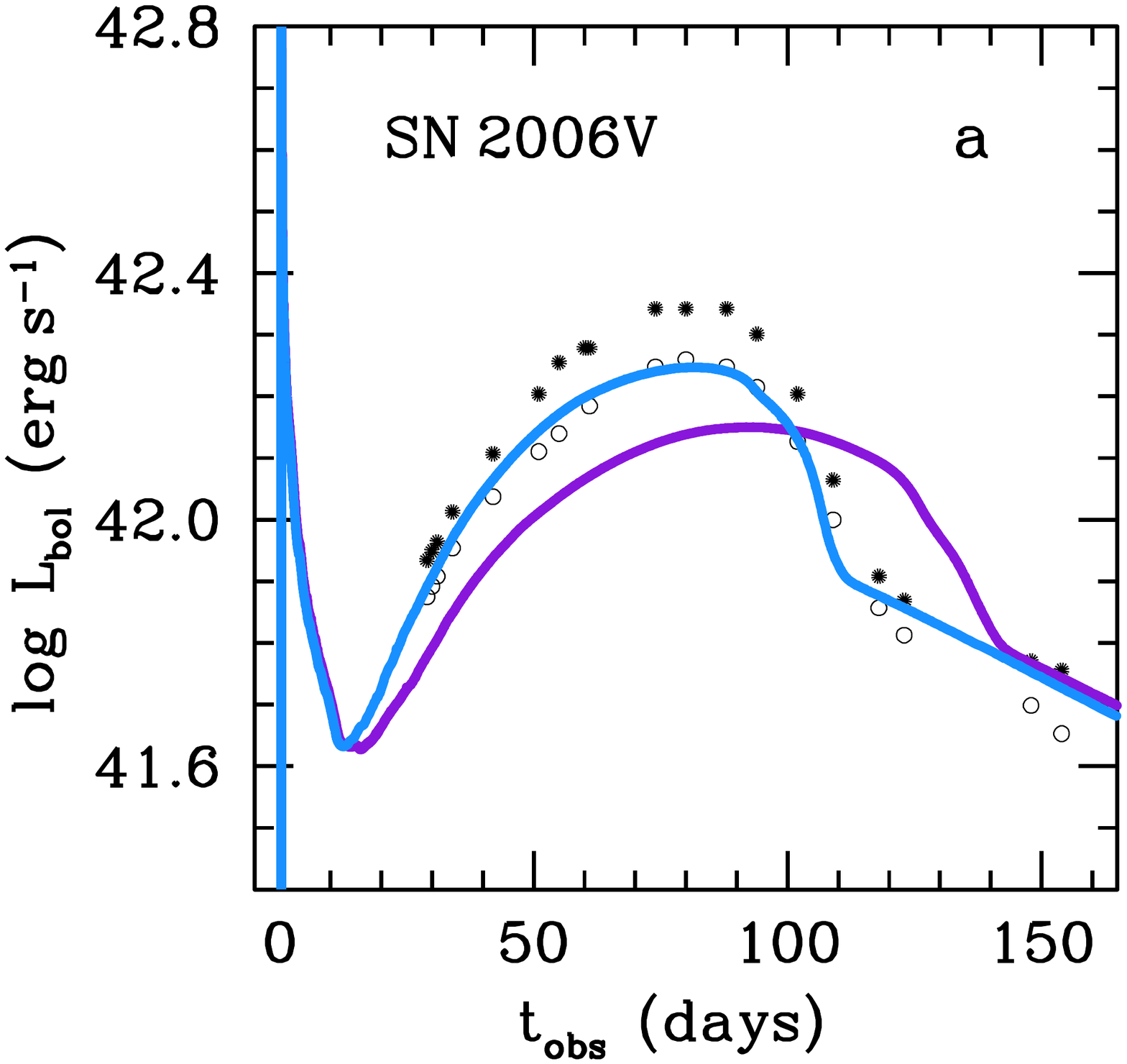}
   \includegraphics[width=0.67\columnwidth, clip, trim=18 182 37 93]{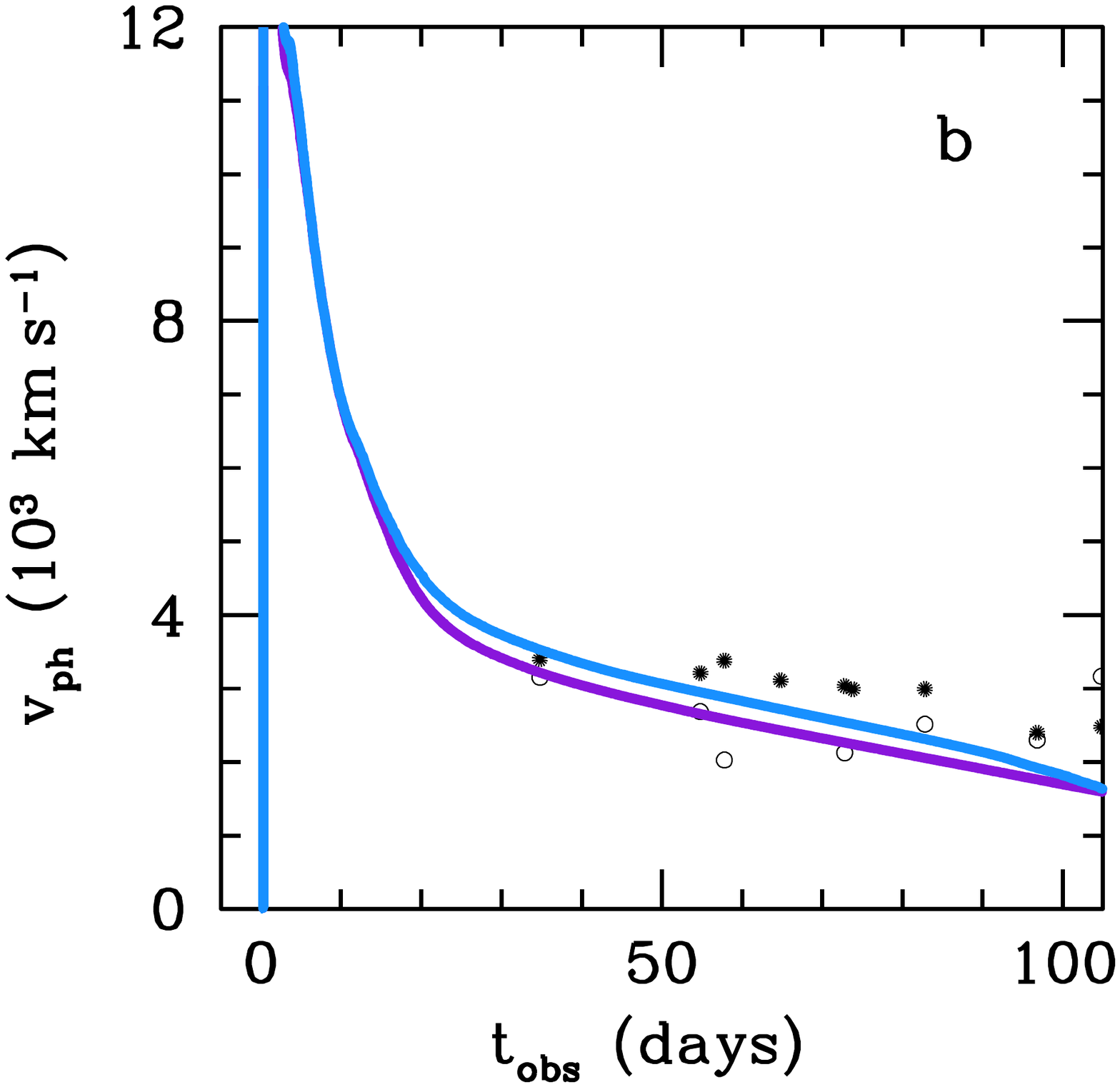}
   \includegraphics[width=0.67\columnwidth, clip, trim=18 182 37 93]{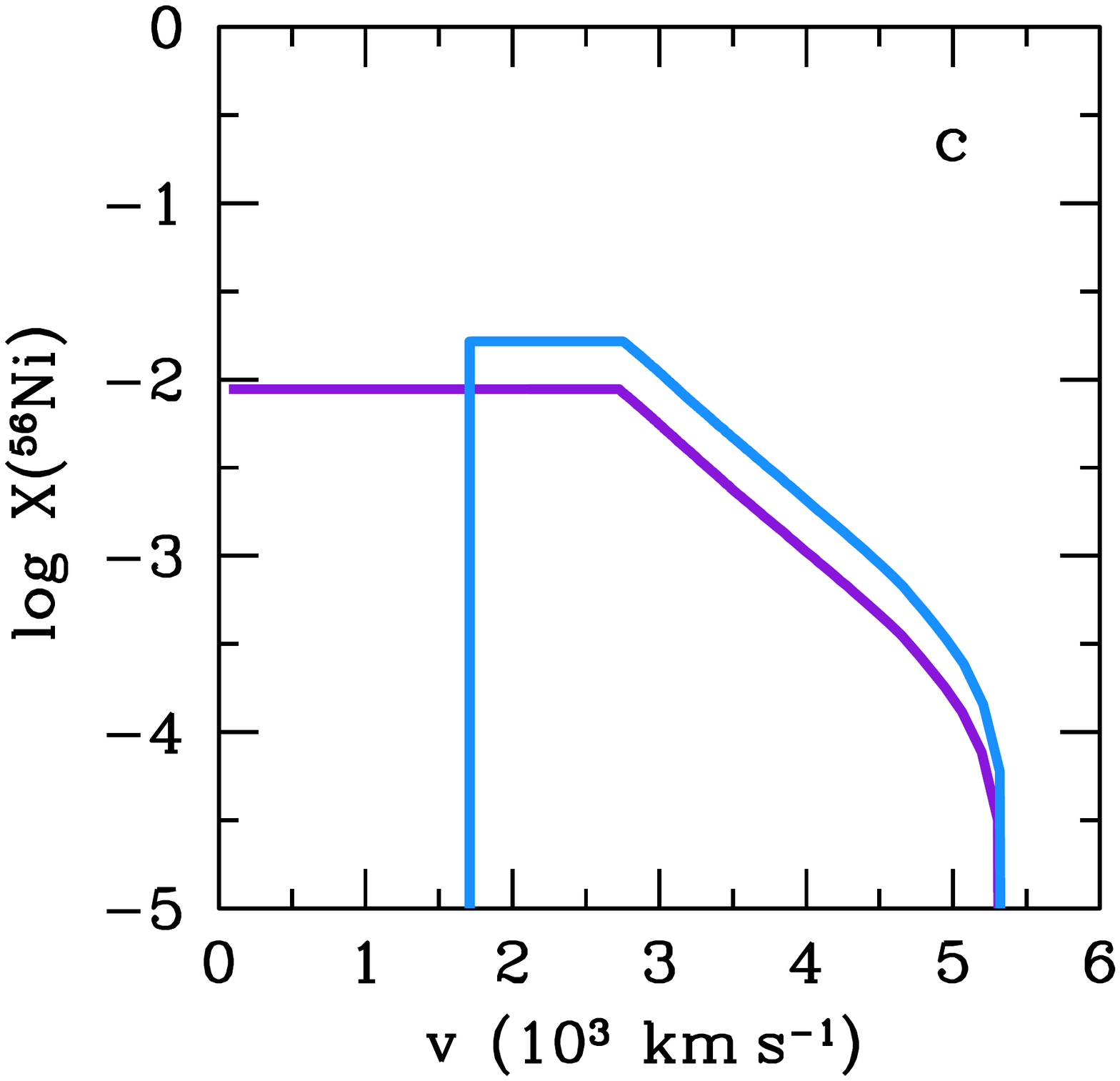}\\
   \caption{
   Hydrodynamic model for SN 2006V.
   \textsl{Panel a}:  Light curves calculated for different $^{56}$Ni distributions
      (see \textsl{Panel c}) are compared with the bolometric data of SN~2006V estimated
      by \citet[][\textsl{open circles}]{taddia2012} and by \citet{lusk2017} which used direct
      integration (\textsl{filled circles}).
   To fit the observations, the explosion date is suggested to be earlier by
      $3\,$days relative to that accepted by \citet{taddia2012}.
   \textsl{Panel b}:  Calculated photospheric velocity is overplotted on the  
      velocity at the photosphere estimated by \citet{taddia2012} with the
      absorption minima of the $\ion{Ba}{ii}~6142\,\AA$ (\textsl{open circles}),
      $\ion{Fe}{ii}~5169$\,\AA\ (\textsl{filled circles}) lines.
   \textsl{Panel c}: Mass fraction of radioactive $^{56}$Ni as a function of
      velocity at  Day 50 for the $^{56}$Ni distributions within sphere
      (\textsl{purple line}) and spherical layer (\textsl{blue line}).
   }
   \label{fig:SN2006V}
\end{figure*}

SN~1998A, which exploded in a spiral arm of the SBc galaxy IC 2627, was a more energetic explosion than SN~1987A \citep{williams1998, woodings1998}, whose photometric data was obtained by \citet{pastorello2005}. Its luminosity exceeds that of SN~1987A at all times and from the luminosity of its decay tail, the nickel mass of this supernova is estimated to be $0.09\,\Msun-0.11\,\Msun$ compared to $0.073\,\Msun$ for SN~1987A \citep{pastorello2005, pastorello2012}.
The high photospheric velocities determined from the absorption minima of $\ion{Ba}{ii}$, $\ion{Fe}{ii}$, and $\ion{Sc}{ii}$ lines indicated that the explosion of SN~1998A was $5-6$ times more energetic than SN~1987A \citep{pastorello2005, pastorello2012}.

From a semi-analytic model that used single star pre-SN models,
   \citet{pastorello2005} predicted a progenitor of radius of $R<86.3\,\Rsun$ and
   an ejecta mass of $M_\mathrm{ej}=22\,\Msun$ for SN~1998A
   (Table~\ref{predicted}) by scaling with progenitor quantities of
   $R<71\,\Rsun$ and $M_\mathrm{ej}=18\,\Msun$ for SN~1987A.
Our results by fitting binary merger models are quite different from these
   predictions. 

Using CRAB, we determine the best fit progenitor model as one that matches both
   the light curve and the photospheric velocities estimated from weak lines
   (Fig.~\ref{fig:SN1998A}).
We test  Model \texttt{16-7b} for SN~1998A, whose density distribution and chemical
   composition averaged by the boxcar mixing with the mass width
   $\Delta{}M=2\,\Msun$ prior to the explosion are shown in
   Fig.~\ref{fig:pSN-16-18}.
Both the luminosity during the outburst of SN~1998A and its photospheric
   velocity (Fig.~\ref{fig:SN1998A}) are remarkably higher than those of
   SN~1987A (Fig.~\ref{fig:SN1987A}).
This indicates a greater explosion energy and a higher nickel mass compared to
   SN~1987A (Table~\ref{tab:hydmod}).
Model \texttt{16-7b} reproduces the light curve shape and nearly matches the
   photospheric velocity profile for $E=4.5\,$B and $M_\mathrm{Ni}=0.12\,\Msun$
  (determined from the luminosity in the radioactive decay tail).
The light curve smoothly rising to the dome maximum requires a strong mixing of nickel up to $5,\!300\,\kms$ to avoid the formation of the local dip in the  luminosity at around Day 20 (Fig.~\ref{fig:SN1998A}\textsl{c}).

SN~2006V is more luminous compared to SN~1987A.  It occurred in the spiral  galaxy UGC 6510 \citep{chen2006} and its photometric data were taken by \citet{taddia2012}. Using semi-analytic modelling and scaling relations, \citet{taddia2012} found that the progenitor had $R<50\,\Rsun-75\,\Rsun$ and $M_\mathrm{ej}=17\,\Rsun-20\,\Rsun$  and exploded with an energy of $2.4\,\B$ and a large nickel mass of $0.127\,\Msun$.
These results are based on the following SN~1987A properties: $R=33\,\Rsun$, $M_\mathrm{ej}=14\,\Msun$, $E=1.1\,$B and $M_\mathrm{Ni}=0.078\,\Msun$.

For hydrodynamic modelling, SN~2006V is a very
   interesting object.
Confronting it with SN~1987A reveals two important facts.
SN~2006V is more luminous than SN~1987A by about 0.4\,dex
   during the  entire outburst (Figures~\ref{fig:SN2006V}\textsl{a} and \ref{fig:SN1987A}\textsl{a}), whereas the photospheric velocities of these SNe at Day 50 are comparable (Figs~\ref{fig:SN2006V}\textsl{b} and \ref{fig:SN1987A}\textsl{b}). The latter implies that SN~2006V and SN~1987A have comparable average velocities of the ejecta.
The luminosity in the interval from nearly Day 10 to Day 30 depends mainly on the average velocity of the ejecta and the pre-SN radius \citep{utrobin2005}. A comparison  of the average velocities leaves  only one way to increase the SN luminosity, namely, to enlarge the initial radius of the pre-SN model compared to that of SN~1987A.

Our best fit binary merger model (Fig.~\ref{fig:SN2006V}) that matches the  light curve and the photospheric velocity is  Model \texttt{18-4d}, which has a radius of $150.4\,\Rsun$ and an ejecta mass of $19.1\,\Msun$, and formed from the merger of a primary of mass $M_1=18\,\Msun$ and secondary of mass $M_2=4\,\Msun$. The explosion was modelled using $E=1\,$B and a nickel mass of $0.15\,\Msun$. Unlike SN~1987A and SN~1998A in which nickel is mixed all the way down to zero velocity (the centre), in SN~2006V nickel is required to be mixed only within a spherical
   layer, between $1,\!700$ and $5,\!400\,\kms$ (Fig.~\ref{fig:SN2006V}\textsl{c}; \textsl{blue line}).

When nickel is distributed within a sphere (Fig.~\ref{fig:SN2006V}\textsl{c}, \textsl{magenta
   line}), this kind of distribution results in a wider luminosity maximum that is shifted to lower luminosities and to later epochs compared to the observed one (Fig.~\ref{fig:SN2006V}\textsl{a}; \textsl{magenta line}). To solve the problem arisen, we have to modify the light curve in a proper way but having available only one free parameter, namely, a distribution of nickel throughout the ejecta.
The maximum velocity of the nickel ejecta is well specified by the
   smoothness of the rising part of the light curve from about  Day 30 to the luminosity maximum. In order to draw the maximum toward earlier epochs and, consequently, toward higher luminosities, we should decrease the optical depth of the ejecta from the layers where gamma rays deposit their energy to the stellar surface. Keeping the maximum velocity of the nickel ejecta, the solution of the problem
   is moving the inner $^{56}$Ni-rich layers out of the centre.
In other words, we should make the central cavity free of radioactive nickel (Fig.~\ref{fig:SN2006V}\textsl{c}; \textsl{blue line}).
This gives a good agreement of the calculated light curve with that observed (Fig.~\ref{fig:SN2006V}\textsl{a}; \textsl{blue line}).

\begin{table*}
\centering
\caption[]{Hydrodynamic models of peculiar Type IIP supernovae.}
\label{tab:hydmod}
\begin{tabular}{lccccccrclcc}
\hline
\noalign{\smallskip}
SN & Model & $M_1$ & $M_2$ & $f_\mathrm{sh}$ & $M_\mathrm{pre-SN}$ &%
 $M_\mathrm{ej}$ & $R_0$ & $E$ & $M_\mathrm{Ni}$ & $v_\mathrm{Ni}^\mathrm{min}$ &%
 $v_\mathrm{Ni}^\mathrm{max}$ \\
   &       & (\Msun) & (\Msun) & (\%) & (\Msun) & (\Msun) &%
   (\Rsun) & ($10^{51}$\,erg) & (\Msun) & (\kms) & (\kms) \\
\noalign{\smallskip}
\hline
\noalign{\smallskip}
1987A & 16-7b & 16 & 7 & 50 & 22.0 & 20.4 & 37.4 & 1.7 & 0.073 & 0   & 3,000 \\
1998A & 16-7b & 16 & 7 & 50 & 22.0 & 20.4 & 37.4 & 4.5 & 0.12   & 0   & 5,300 \\
2006V & 18-4d & 18 & 4 & 90 & 20.5 & 18.8 & 150.3 & 1.0 & 0.15 & 1700 & 5,400 \\
\noalign{\smallskip}
\hline
\end{tabular}
\end{table*}

%% file: Conclusions.tex
\section{Discussions and Conclusions}
\label{conclusions}

In this paper, we have presented the results of the first explosion study of progenitors from binary mergers for peculiar Type IIP SNe.
Our binary merger models for SN~1987A are significantly different from the evolutionary single star progenitor models of \citet{woosley1988a,
   shigeyama1990, woosley1997,woosley2007a}: they have smaller He-core masses
   ($3\,\Msun-4\,\Msun$ compared to $4\,\Msun-7.4\,\Msun$), larger envelope masses ($16\,\Msun-19.2\,\Msun$
   compared to $9.5\,\Msun-13.6\,\Msun$) and smaller radii ($29\,\Rsun-64\,\Rsun$ compared to
   $47\,\Rsun-64\,\Rsun$).
The overall compact structure and large envelope mass of the merger progenitor
   model are close to the characteristics predicted by an ``optimal'' non-evolutionary
   model of \citet{utrobin2005}, which has $R=35\,\Rsun$ and $M_\mathrm{ej}=18\,\Msun$.
The binary merger models also have a steeper density gradient at the
   He-core/H-envelope interface and a smaller compactness parameter compared
   to single star models.

From the set of progenitor models we explored, the explosion of Model \texttt{16-7b} from the merger of a primary star of mass $M_1=16\,\Msun$, secondary of mass $M_2=7\,\Msun$, with a dredge-up efficiency of $f_\mathrm{sh}=50\,\%$,  resulting in a pre-supernova model which has
   a radius of $R=37.4\Rsun$, a helium core mass of $M_\mathrm{He,c}=3.4\Msun$, and  an ejecta mass of  $M_\mathrm{ej}=20.6\Msun$, exploding
   with an energy of $1.7\,$B, nickel mixing velocity of $3,\!000\,\kms$ and nickel mass of $0.073\,\Msun$ matches the light curve and photospheric velocity profile of SN~1987A (Fig.~\ref{fig:SN1987A}).  
The fit of this model to the light curve data is a significant improvement over current explosion models from single stars, especially in matching the luminosity dip at Day $\sim$8 and the dome shape between Days $40-120$ (Fig.~9 in \citealp{utrobin2015}).
Model \texttt{16-7b} also satisfies the observational constraints of the progenitor \sk \citep{menon2017}, with $\teff=15.8\,$kK, $\log\,L/\mathrm{L}_{\odot}=4.9$, and $\text{N/C}=6.9$, $\text{N/O}=1.4$,
   and $\text{He/H}=0.14$, making it the first evolutionary model in literature that is compatible with the observations of the progenitor and the explosion properties of SN~1987A. The explosion parameters that produce the best fit to the light curve and
   photospheric velocity of SN~1987A are: $E=1.7\,\B$, nickel mass of $0.073\,\Msun$ and nickel mixing velocity up to $3,\!000\,\kms$ which agrees with spectroscopic measurements.
These results are in agreement with the predictions of \citet{utrobin2004, utrobin2005} from their studies of non-evolutionary models.

The ejecta mass we obtain for the pre-SN Model \texttt{16-7b} on including a part of the C-O core, contains $M(\text{O})=1.23\,\Msun$, $M(\text{He})=7.14\,\Msun$, and $M(\text{H})=11.50\,\Msun$.  Studies of the nebular phase spectra of SN~1987A give mass estimates of these species.  Oxygen, which is the most abundant of the metals and a good  probe of the progenitor mass for single star models, contributes $1.2\,\Msun-2.0\,\Msun$  \citep{chugai1994, kozma1998, chugai1997}.  The masses of helium and hydrogen are $\sim5.8\,\Msun$ and $\sim3.9\,\Msun$, respectively \citep{kozma1998}.  The model values of oxygen and helium are comparable to the spectral values, whereas the hydrogen mass of the ejecta is larger than its observational estimate by a factor of three.  We have no ready explanation for the disparity found between the hydrogen masses.  We can, however, make a general remark that hydrodynamic modeling of the observed light curves from the onset of the explosion to the radioactive tail involves all the ejected matter and gives  better estimates of the total element masses than exploring the emission lines observed at the nebular phase\, which provides the element abundances only in the restricted parts, not in the whole ejecta.

The choice of initial parameters for the merger models, i.e., the fraction of the He shell of the primary's He core dredged up ($f_\mathrm{sh}$), the mass of the secondary star ($M_2$) and the mass of the primary star ($M_1$), which together determine the mass of the He core, envelope mass
   and the radius of the pre-SN star, also affect the light curve shape.
For a given $M_1$ and $M_2$, increasing $f_\mathrm{sh}$ causes the He-core
   mass to decrease and the structure to become radially more extended.
The explosion of progenitor models with decreasing He-core mass affects
   the luminosity dip at around Day 8, causing the light curve to ascend more
   rapidly and shifting it upward overall.
On the other hand, increasing $M_2$ for given $M_1$ and $f_\mathrm{sh}$
   causes the pre-SN model's envelope mass to increase and its radius to decrease.
The explosion of these models causes the light curve to descend further
   at Day $\sim$8 and to have a more pronounced delayed rise to the luminosity maximum between Days $15-40$. Consequently, the light curve moves further to the right as $M_2$ increases, keeping nearly the same characteristic width of the dome.

We extended the study to investigate the explosions of our pre-SN models for two other peculiar Type IIP SNe, SN~1998A and SN~2006V. The explosion of Model \texttt{16-7b}, with $E=4.5\,$B, a nickel mass of $0.12\,\Msun$ and
 a nickel mixing velocity up to $5,\!300\,\kms$ reproduced the light curve shape and reasonably matched the photospheric velocity evolution of SN~1998A. Using a semi-analytic model, \citet{pastorello2005} predicted the progenitor
   for this supernova to have had a radius less than $86.3\,\Rsun$ and ejecta mass
   of $22\,\Msun$ and to have exploded with an energy of $5\,\B-6\,$B.
Thus our results are in reasonable agreement with the predictions of
   \citet{pastorello2005}.

According to our study, SN~2006V had a much more extended progenitor, with
   $R=150.3\,\Rsun$ and $M_\mathrm{ej}=19.1\,\Msun$.
This pre-SN model was created from the merger of a primary of $M_1=18\,\Msun$,
   $M_2=4\,\Msun$ with $f_\mathrm{sh}=90\,\%$, and had a He-core mass of $3.8\,\Msun$,
   an envelope mass of $16.7\,\Msun$, $\teff=7.5\,$kK, and a luminosity of $\log\,L/\mathrm{L}_{\odot}=4.8$.
Our progenitor for this supernova thus is not a blue supergiant, but a yellow supergiant.
On the other hand, \citet{taddia2012}, who did the photometric analysis of this supernova, predicted a progenitor with $R<50\,\Rsun-75\,\Rsun$ and
   $M_\mathrm{ej}=17\,\Msun-20\,\Msun$, based on a semi-analytic model.
Thus our results for the progenitor differ significantly from those predicted
   by \citet{taddia2012}.
The corresponding explosion parameters in our study are: $E=1\,\B$, nickel
   mass of $0.15\,\Msun$ and nickel mixing in the regions with velocities
   between $1,\!700-5,\!400\,\kms$.
The latter result is different from the nickel mixing velocity profiles
   for SN~1987A and SN~1998A in which the nickel was mixed down all the way
   to the centre where the velocity was zero.
It is remarkable that this kind of $^{56}$Ni distribution with the central
   cavity free of nickel is similar to that derived for the luminous Type IIP
   SN~2013ej \citep{UC_17}.
Again, our explosion parameters are different from \citet{taddia2012} who
   calculated an explosion energy of $2.4\,$B and a nickel mass of $0.127\,\Msun$.
These results of \citet{taddia2012} may have to do with the progenitor model
   assumed for SN~1987A, which had a much smaller ejecta than our progenitor
   model, with $M_\mathrm{ej}=14\,\Msun$ compared to $M_\mathrm{ej}=20.6\,\Msun$
   and a smaller explosion energy of $E=1.1\,$B as compared to $E=1.7\,$B.

The light curve of SN~2000cb had a much broader dome than that of SN~1987A
   and other peculiar Type IIP SNe and was a more energetic explosion than
   SN~1987A, with $E=4\,$B \citep{kleiser2011} and $E=4.4\,$B
   \citep{utrobin2011}.
The ``optimal'' non-evolutionary progenitor model for SN~2000cb calculated by
   \citet{utrobin2011} has $R=35\pm14\,\Rsun$ and $M_\mathrm{ej}=22.3\pm\,1\Msun$.
Although our merger progenitor models can match these constraints on radius
   and ejecta mass, they are structurally not suitable to reproduce the
   unusual light curve shape of this supernova.
We also could not study SN~2006au due to data missing from the first $50\,$days of the supernova and  neither SN~2009E because its nickel decay tail in the light curve was not recorded.

Studying the mixing of nickel and hydrogen is promising with these new pre-supernova models.  As BSG progenitors may not only form from the late Case C merger scenario developed in this work, it would also be interesting to study the core-envelope structure of a BSG produced from a Case B merger/accretion scenarios and the results of their explosion.  Currently, 3D explosion simulations of our models are underway along with a separate study of the nuclear yields from their explosions.  